\documentclass[prc,aps,twocolumn,showpacs]{revtex4}
\usepackage{graphicx}
\def\pcomma{$^,$}
\def\pglsg{$^1$}
\def\pucla{$^{2}$}
\def\pgwu{$^3$}
\def\pmainz{$^4$}
\def\pgatch{$^5$}
\def\pkent{$^6$}
\def\pbonn{$^7$}
\def\pgiess{$^8$}
\def\ppavia{$^9$}
\def\pedinb{$^{10}$}
\def\plpi{$^{11}$}
\def\ptom{$^{12}$}
\def\psackv{$^{13}$}
\def\pbasel{$^{14}$}
\def\pinr{$^{15}$}
\def\pzagreb{$^{16}$}
\def\pcua{$^{17}$}

\begin{document}

\title{Study of the $\gamma p\to\eta p$ reaction with 
	the Crystal Ball detector at the Mainz Microtron 
	(MAMI-C)}

\author{E.~F.~McNicoll\pglsg,
        S.~Prakhov\pucla\footnote[2]{Electronic address: prakhov@ucla.edu},
        I.~I.~Strakovsky\pgwu\footnote[4]{Electronic address: igor@gwu.edu},
	P.~Aguar-Bartolom\'e\pmainz,
        L.~K.~Akasoy\pmainz,
        J.~R.~M.~Annand\pglsg,
        H.~J.~Arends\pmainz,
        R.~A.~Arndt\pgwu\footnote[1]{Deceased},
        Ya.~I.~Azimov\pgatch,
        K.~Bantawa\pkent,
        R.~Beck\pbonn,
        V.~S.~Bekrenev\pgatch,
        H.~Bergh\"auser\pgiess,
        A.~Braghieri\ppavia,
        D.~Branford\pedinb,
        W.~J.~Briscoe\pgwu,
        J.~Brudvik\pucla,
        S.~Cherepnya\plpi,
	R.~F.~B.~Codling\pglsg,
	B.~T.~Demissie\pgwu,
	E.~J.~Downie\pmainz\pcomma\pglsg\pcomma\pgwu,
	P.~Drexler\pgiess,
	L.~V.~Fil'kov\plpi,
	A.~Fix\ptom,
	B.~Freehart\pgwu,
	D.~I.~Glazier\pedinb,
	R.~Gregor\pgiess,
	D.~Hamilton\pglsg,
	E.~Heid\pmainz\pcomma\pgwu,
	D.~Hornidge\psackv,
	I.~Jaegle\pbasel,
	O.~Jahn\pmainz,
	T.~C.~Jude\pedinb,
	V.~L.~Kashevarov\plpi,
	J.~D.~Kellie\pglsg,
	I.~Keshelashvili\pbasel,
	R.~Kondratiev\pinr,
	M.~Korolija\pzagreb,
	M.~Kotulla\pgiess,
	A.~A.~Koulbardis\pgatch,
	S.~P.~Kruglov\pgatch,
	B.~Krusche\pbasel,
	V.~Lisin\pinr,
	K.~Livingston\pglsg,
	I.~J.~D.~MacGregor\pglsg,
	Y.~Maghrbi\pbasel,
	D.~M.~Manley\pkent,
	Z.~Marinides\pgwu,
	M.~Martinez\pmainz,
	J.~C.~McGeorge\pglsg,
	B.~McKinnon\pglsg,
	D.~Mekterovic\pzagreb,
	V.~Metag\pgiess,
	S.~Micanovic\pzagreb,
	D.~Middleton\psackv,
	A.~Mushkarenkov\ppavia,
	B.~M.~K.~Nefkens\pucla,
	A.~Nikolaev\pbonn,
	R.~Novotny\pgiess,
	M.~Ostrick\pmainz,
	P.~B.~Otte\pmainz,
	B.~Oussena\pmainz\pcomma\pgwu,
	M.~W.~Paris\pgwu,
	P.~Pedroni\ppavia,
	F.~Pheron\pbasel,
	A.~Polonski\pinr,
	J.~Robinson\pglsg,
	G.~Rosner\pglsg,
	T.~Rostomyan\ppavia\footnote[3]{Present address:
           University of Basel, Switzerland},
	S.~Schumann\pmainz,
	M.~H.~Sikora\pedinb,
	D.~I.~Sober\pcua,
	A.~Starostin\pucla,
	I.~M.~Suarez\pucla,
	I.~Supek\pzagreb,
	M.~Thiel\pgiess,
	A.~Thomas\pmainz,
	L.~Tiator\pmainz,
	M.~Unverzagt\pmainz,
	D.~P.~Watts\pedinb,
	D.~Werthm\"uller\pbasel,
	R.~L.~Workman\pgwu,
	I.~Zamboni\pzagreb, and
	F.~Zehr\pbasel\\
\vspace*{0.1in}
(Crystal Ball Collaboration at MAMI)
\vspace*{0.1in}
}
\affiliation{
\pglsg Department of Physics and Astronomy, University
        of Glasgow, Glasgow G12 8QQ, United Kingdom}
\affiliation{
\pucla University of California Los Angeles, Los Angeles,
    California 90095-1547, USA}
\affiliation{
\pgwu The George Washington University, Washington, DC
        20052, USA}
\affiliation{
\pmainz Institut f\"ur Kernphysik, University of Mainz,
        D-55099 Mainz, Germany}
\affiliation{
\pgatch Petersburg Nuclear Physics Institute, 188300
        Gatchina, Russia}
\affiliation{
\pkent Kent State University, Kent, Ohio 44242, USA}
\affiliation{
\pbonn Helmholtz-Institut f\"ur Strahlen- und Kernphysik,
        University of Bonn, D-53115 Bonn, Germany}
\affiliation{
\pgiess II Physikalisches Institut, University of Giessen,
        D-35392 Giessen, Germany}
\affiliation{
\ppavia INFN Sezione di Pavia, I-27100 Pavia, Italy}
\affiliation{
\pedinb School of Physics, University of Edinburgh, 
	Edinburgh EH9 3JZ, United Kingdom}
\affiliation{
\plpi Lebedev Physical Institute, 119991 Moscow, Russia}
\affiliation{
\ptom Tomsk Polytechnic University, 634050 Tomsk, Russia}
\affiliation{
\psackv Mount Allison University, Sackville, New Brunswick
    E4L3B5, Canada}
\affiliation{
\pbasel Institut f\"ur Physik, University of Basel,
        CH-4056 Basel, Switzerland}
\affiliation{
\pinr Institute for Nuclear Research, 125047 Moscow, 
	Russia}
\affiliation{
\pzagreb Rudjer Boskovic Institute, HR-10000 Zagreb, 
	Croatia}
\affiliation{
\pcua The Catholic University of America, Washington,
    DC 20064, USA}

\date{\today}

%%%%%%%%%%%%%%%%%%%%%%%%%%%%%%%%%%%%%%%%%%%%%%%%%%%%%%%
\begin{abstract}
The $\gamma p\to\eta p$ reaction has been measured with
the Crystal Ball and TAPS multiphoton spectrometers
in the energy range from the production
threshold of 707~MeV to 1.4~GeV ($1.49\le
W\le 1.87$~GeV).  Bremsstrahlung photons produced by
the 1.5-GeV electron beam of the Mainz Microtron MAMI-C
and momentum analyzed by the Glasgow Tagging Spectrometer
were used for the $\eta$-meson production.  Our 
accumulation of $3.8\times 10^6$ $\gamma p\to\eta p\to 
3\pi^0p\to 6\gamma p$ events allows a detailed study of 
the reaction dynamics. The $\gamma p\to\eta p$ 
differential cross sections were determined for 120 
energy bins and the full range of the production angles.  
Our data show a dip near $W=1680$~MeV in the total 
cross section caused by a substantial dip in $\eta$ 
production at forward angles. The data are compared
to predictions of previous SAID and MAID partial-wave
analyses and to the latest SAID and MAID fits that
included our data.
\end{abstract}


\maketitle

%%%%%%%%%%%%%%%%%%%%%%%%%%%%%%%%%%%%%%%%%%%%%%%%%%%%%%
\section{Introduction}
\label{sec:intro}

The $N^\ast$ family of nucleon resonances has many well
established members~\cite{PDG}, several of which exhibit
overlapping pole positions, very similar masses and 
widths, but different $J^P$ spin-parity values.  Apart 
from the $N(1535)1/2^-$ state, the known photo-decay
amplitudes have been determined from analyses of 
single-pion photoproduction, so the $\eta N$ branching 
ratios are in general poorly known. Evidently 
$\eta$-photoproduction data are required, and this work 
studies the region from threshold, where there are two 
closely spaced states: $N(1520)3/2^-$ and $N(1535)1/2^-$, 
up to center-of-mass (c.m.) energies of $W\approx 
1800$~MeV, encompassing a sequence of six overlapping 
states: $N(1650)1/2^-$, $N(1675)5/2^-$, $N(1680)5/2^+$,
$N(1700)3/2^-$, $N(1710)1/2^+$, and $N(1720)3/2^+$.
Compared to pion photoproduction, the $\eta$ channel
has some advantages. Isospin conservation requires
that $\eta$ production probes only $I=1/2$ 
contributions, simplifying the extraction of individual 
$N^\ast$ properties. 
%Unitary, $\eta N$-threshold cusp 
%effects, which distort the $S$-wave pion amplitudes, 
%are avoided as the $\eta$ channel is dominated by the 
%$N(1535)1/2^-$ state.
Couplings of $N^\ast$s to the $\eta N$ channel, in 
comparison with the $\pi N$ couplings, may clarify 
the inner structure of resonances.

New, high quality data on $\gamma p\to\eta p$ are 
needed to shed light on these issues, and the 
tagged-photon hall at Mainz offers a state-of-the-art 
facility to obtain such data. Here we report on a new 
differential-cross-section measurement, covering incident 
photon energies from threshold ($E_\gamma = 707$~MeV) up 
to $E_\gamma = 1400$~MeV.  The accumulation of $3.8\times 
10^6$ events for the process $\gamma p\to\eta p\to 
3\pi^0p\to 6\gamma p$ has enabled the data to be binned 
finely in $E_\gamma$ (bin widths as small as $\sim 4$~MeV) 
and in $\eta$ production angle, which in the c.m. frame 
is fully covered.  The present measurement is part of an 
extensive program at the Mainz Microtron to provide 
data of unrivaled quality on neutral meson photoproduction, 
which includes polarized beam and target observables in 
addition to cross sections.

Our energy range includes several well-established
resonances and also some more questionable ones.
Indeed, the excellent photon-energy resolution offers
the potential to illuminate any narrow states, possibly 
of exotic structure. Most of the states presently 
covered appear to have very small coupling to the $\eta 
N$ channel, and this in itself can be puzzling.  For 
example, it is unclear why the $\eta N$ branching 
ratio is so small for the second $S_{11}$, 
$N(1650)1/2^-$, compared to the first $N(1535)1/2^-$. 
The data available for $\pi^-p\to\eta n$ are
inadequate to study this question~\cite{sp06,nstar09}.  
The reason for a small branching ratio of $N(1520)3/2^-$ 
to $\eta N$~\cite{PDG,ar05} has to be understood, too.  
The Particle-Data-Group (PDG) estimate for the 
A$_{3/2}$ decay amplitude of the $N(1720)3/2^+$ state
is consistent with zero, while the recent SAID 
determination gives a small but non-vanishing
value~\cite{du09}. The reason for the disagreement 
between the PDG estimate for the A$_{1/2}$ decay 
amplitude and the recent SAID determination
\cite{du09} is also unclear. Other unresolved issues
relate to the second $P_{11}$ and $D_{13}$ 
resonances [$N(1710)1/2^+$ and $N(1700)3/2^-$] that 
are not seen in the recent $\pi N$ partial-wave 
analysis (PWA)~\cite{sp06}, contrary to other PWAs
used by the Particle Data Group~\cite{PDG}. 
The $\eta N$ decay channel could be more favorable 
than $\pi N$ for these states.  The present data 
should have sufficient precision to allow reliable 
extraction of the $\eta N$ partial waves for these 
resonances, which will enhance our understanding 
of their internal dynamics. In addition, since the 
present data have good coverage of the $\eta 
p$-threshold region, the S-wave dominance of the 
threshold behavior can also be checked.

The paper is laid out in the following manner: the
experimental setup is briefly described in
Sec.~\ref{sec:Exp}; the procedure to determine the
differential cross sections is described in
Sec.~\ref{sec:Data}; the estimation of our
systematic uncertainties is given in
Sec.~\ref{sec:Errs}; the experimental results are
presented in Sec.~\ref{sec:Results}; analyses
of the data in terms of SAID and MAID
are described in Sec.~\ref{sec:Fit};
finally, the findings of our study are summarized
in Sec.~\ref{sec:Conc}.

Since the present data on $\gamma p\to\eta p\to 3\pi^0p$
were also used in the determination of the slope
parameter $\alpha$ for the $\eta\to 3\pi^0$
decay~\cite{slopemamic}, a more detailed description
of the experiment and data handling can be found in 
Ref.~\cite{slopemamic}.

%%%%%%%%%%%%%%%%%%%%%%%%%%%%%%%%%%%%%%%%%%%%%%%%%%
\section{Experimental setup}
\label{sec:Exp}

The process $\gamma p\to\eta p\to 3\pi^0p\to 6\gamma
p$ was measured using the Crystal Ball (CB)~\cite{etalam}
as the central spectrometer and the Two-Arm Photon 
Spectrometer (TAPS)~\cite{TAPS,TAPS2}
as a forward spectrometer. These detectors were
installed in the tagged bremsstrahlung photon beam of
the Mainz Microtron (MAMI)~\cite{MAMI,MAMIC}, with the
photon energies determined by the Glasgow Tagging
Spectrometer~\cite{Anthony,Hall,Tagger2}.

The CB spectrometer is a sphere consisting of 672
optically insulated NaI(Tl) crystals, shaped as
truncated triangular pyramids, which point toward
the center of the sphere. Each NaI(Tl) crystal is 
41~cm long, which corresponds to 15.7 radiation 
lengths.  The crystals are arranged in two 
hemispheres that cover 93\% of $4\pi$ sr, sitting 
outside a central spherical cavity with a radius of 
25~cm, which is designed to hold the target and inner 
detectors.  To allow passage of the beam, the regions 
of polar angle below $20^\circ$ and above $160^\circ$ 
are not populated.  The energy resolution for 
electromagnetic showers in the CB can be described as 
$\Delta E/E = 0.020/(E[\mathrm{GeV}])^{0.36}$.  
Shower directions are determined with a resolution in 
$\theta$, the polar angle with respect to the beam 
axis, of $\sigma_\theta = 2^\circ \textrm{--} 
3^\circ$, under the assumption that the photons are 
produced in the center of the CB. The resolution in 
the azimuthal angle $\phi$ is $\sigma_\theta/\sin\theta$.

To cover the forward aperture of the CB, the TAPS 
calorimeter~\cite{TAPS,TAPS2} was installed 1.5~m 
downstream of the CB center.  TAPS geometry is 
flexible and, for the present A2 experiment, it was 
configured as a ``plug" for the forward-angle hole 
in CB acceptance.  In this experiment, TAPS was 
arranged in a plane consisting of 384 BaF$_2$ 
counters of hexagonal cross section, with an inner 
diameter of 5.9~cm and a length of 25~cm, which 
corresponds to 12 radiation lengths.  One counter 
was removed from the center of the array, which has 
an over-all, hexagonal geometry in the $x$-$y$ plane, 
to allow the passage of the photon beam.  TAPS 
subtends the full azimuthal range for polar angles
from $1^\circ$ to $20^\circ$.  The energy 
resolution for electromagnetic showers in the TAPS 
calorimeter can be described as $\Delta E/E = 
0.018 + 0.008/(E[\mathrm{GeV}])^{0.5}$.  Because 
of the relatively long distance from the CB, the 
resolution of TAPS in the polar angle $\theta$ 
was better than $1^\circ$. The resolution of TAPS 
in the azimuthal angle $\phi$ is better than $1/R$
radian, where $R$ is the distance in centimeters
from the TAPS center to the point on the TAPS
surface that corresponds to the $\theta$ angle.

The upgraded Mainz Microtron, MAMI-C, is a four 
stage accelerator, and its latest addition (the 
fourth stage) is a harmonic double-sided electron 
accelerator~\cite{MAMIC}.
An electron-beam energy of 1508 MeV was used
for the present experiment.
Bremsstrahlung photons, produced by electrons 
in a 10-$\mu$m Cu radiator and collimated by a 
4-mm-diameter Pb collimator, were incident on a 
5-cm-long liquid hydrogen (LH$_2$) target located 
in the center of the CB. The energies of the incident 
photons were analyzed up to 1402~MeV by detecting 
the post-bremsstrahlung electrons in the Glasgow 
Tagger~\cite{Anthony,Hall,Tagger2}.  The Tagger is 
a broad-momentum-band, magnetic-dipole spectrometer 
that focuses post-bremsstrahlung electrons onto a 
focal-plane detector, consisting of 353 half-overlapping 
plastic scintillators. The energy resolution of 
the tagged photons is mostly defined by the
overlap region of two adjacent scintillation 
counters (a tagger channel) and the electron beam 
energy. For a beam energy of 1508~MeV, a tagger 
channel has a width about 2~MeV at 1402~MeV and 
about 4~MeV at 707~MeV (the $\eta$-production 
threshold).  Tagged photons are selected in the 
analysis by examining the correlation in time 
between a tagger channel and the experimental 
trigger derived from CB signals.

The LH$_2$ target is surrounded by a particle
identification (PID) detector~\cite{PID} which 
is a cylinder of length 50 cm and diameter 
12~cm, built from 24 identical plastic 
scintillator segments, of thickness 0.4~cm. In 
conjunction with the CB, this identifies 
charged particles by the $\Delta E/E$ 
technique, although this facility was not used 
in the present analysis.

The experimental trigger had two main 
requirements. First, the sum of the pulse
amplitudes from the CB crystals had to exceed 
a hardware threshold that corresponded to an
energy deposit of $\sim 320$~MeV. Second, the 
number of ``hardware" clusters in the CB had 
to be larger than 2. A ``hardware" cluster 
is a group of 16 adjacent crystals in which at 
least one crystal has an energy deposit larger 
than 30~MeV.

%%%%%%%%%%%%%%%%%%%%%%%%%%%%%%%%%%%%%%%%%%%
\section{Data Handling}
\label{sec:Data}

The photoproduction of $\eta$ was measured by using 
the $3\pi^0$ decay mode of this meson:
\begin{equation}
    \gamma p\to\eta p\to 3\pi^0p\to 6\gamma p.
    \label{eqn:eta3pi0}
\end{equation}
The other main neutral decay mode, $\eta\to\gamma\gamma$,
was not used in this measurement because a large
number of $\eta\to\gamma\gamma$ events did not
satisfy the experimental-trigger requirements.
The acceptance determination for the two-photon
final state then becomes highly sensitive to the 
trigger efficiency, calculated using a Monte Carlo 
(MC) simulation.  On the contrary, the trigger 
efficiency for $\eta\to 3\pi^0$ events is close to 
100\%, so the systematic uncertainty in the 
acceptance caused by the trigger simulation is 
small. Additionally, the angular resolution for 
$\eta\to\gamma\gamma$ events is worse than for 
$\eta\to 3\pi^0$, and there is a substantial 
background from the $\gamma p\to\pi^0p$ reaction, 
which requires a careful subtraction.

Process~(\ref{eqn:eta3pi0}) was investigated by 
analysis of events having six and seven 
``software" clusters reconstructed in both the 
CB and TAPS.  The six-cluster sample was used 
to search for the events in which only six 
photons were detected, while for seven clusters
the recoiling proton was also detected. 

\begin{figure}
\includegraphics[width=8.cm,height=5.7cm,bbllx=1.cm,bblly=0.5cm,bburx=19.2cm,bbury=12.cm]{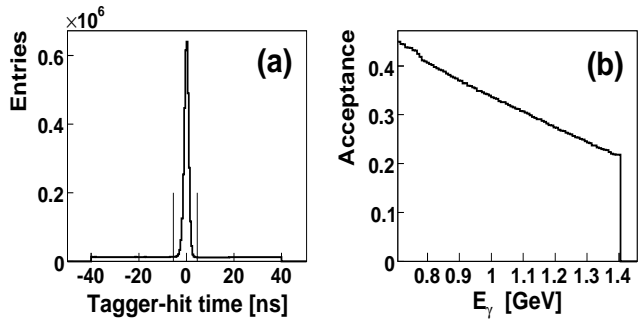}
\caption{(a) Tagger-hit time distribution for the
    experimental $\gamma p\to\eta p\to 3\pi^0p$ events,
    where the two vertical lines define the region of
    prompt coincidences.
    (b) Detector acceptance for $\gamma p\to\eta p\to 3\pi^0p$ 
    events as a function of the incident-photon 
    energy, $E_\gamma$.}\label{fig:beameta_mamic}
\end{figure}

The kinematic-fitting technique was used to test
the reaction hypotheses needed in our analysis 
and to select good candidates for the events of 
interest.  The details of our parametrization of 
the detector information and resolutions are 
given in Ref.~\cite{slopemamic}. The events that 
satisfied the hypothesis of reaction 
(\ref{eqn:eta3pi0}) at the 2\% confidence level, 
CL, (i.e., with a probability of misinterpretation
less than 2\%) were accepted as $\eta\to 3\pi^0$ 
candidates. The kinematic-fit output was then used 
to reconstruct the kinematics of the reaction. 
Since each event in general included several 
tagger hits (due to the high rates in the tagger
detector), the $\gamma p\to\eta p\to 3\pi^0p\to 
6\gamma p$ hypothesis was tested for each tagger 
hit. Selection was based on the hit time which was 
required to be within a selected window (detailed 
below) and the equivalent photon energy, which was 
required to be above the reaction threshold of 
707~MeV. The tagger-hit time distribution for 
$\eta\to 3\pi^0$ event candidates is shown in 
Fig.~\ref{fig:beameta_mamic}(a). If an $\eta\to 
3\pi^0$ event candidate from one trigger passed 
the 2\%~CL criterion for several tagger hits, they 
were analyzed as separate events. The width of the 
tagger-hit window was chosen to be 
substantially wider (80~ns for this analysis)
than the peak caused by prompt 
coincidences between the tagger and trigger.
The width of the prompt window,
denoted by vertical lines in 
Fig.~\ref{fig:beameta_mamic}(a),
was taken to be 10~ns in order to include all
prompt events.
Using a wider window for the random coincidences
allowed the collection of a sufficient number of 
events to determine precisely the 
random-background distribution beneath the prompt 
peak. The experimental distributions analyzed 
with the pure random events were then used to 
subtract the random background from the 
prompt-plus-random event sample.
For our experimental conditions and for the
chosen tagger-hit window, 40\% of all
event candidates were selected for more than
one tagger hit. Since, for an event, there can be only
one prompt tagger hit with the proper $E_\gamma$, there is no
double counting of good events in the distributions with
prompt candidates. 
%%%%%%%%%%%%%%%%%%%%%%%%%%%%%%%%%%%%%%%%%%%%%%
%%%%%%%%%%%%%%%%%%%%%%%%%%%%%%%%%%%%%%%%%

The Monte Carlo simulation of the $\gamma 
p\to\eta p\to 3\pi^0p$ reaction that was used 
for the determination of the differential 
cross sections assumed an isotropic distribution 
of the production angle and independence of 
the reaction yield on the incident-photon energy. 
The simulation of the $\eta\to 3\pi^0$ decay was 
made according to phase space. The small deviation 
of the actual $\eta\to 3\pi^0$ decays from phase 
space was not significant in our analysis. All 
MC events were propagated through a {\sc GEANT} 
(version 3.21) simulation of the CB-TAPS 
detector, folded with resolutions of the detectors 
and conditions of the trigger. The resultant 
simulated data were analyzed in the same way as 
the experimental data. The resulting detector 
acceptance for the $\gamma p\to\eta p\to 3\pi^0p$ 
events selected by the kinematic fit at the 
2\%~CL is shown in Fig.~\ref{fig:beameta_mamic}(b)
as a function of the incident-photon energy, $E_\gamma$. 
It varies from about 45\% at the $\eta$ threshold 
to about 25\% at an $E_\gamma$ of 
1.4~GeV.  The agreement between the various 
experimental spectra and the spectra from the MC 
simulation has been illustrated in 
Ref.~\cite{slopemamic}.

Besides the random-coincidence background, there 
are two more background sources. The first one 
comes from interactions of incident photons with 
the target walls, which was investigated by
analyzing the data taken when the target was 
empty.  It was determined that the fraction of 
the empty-target background that remained in 
our $\eta\to 3\pi^0$ event candidates after 
applying the 2\%-CL cut varied from 1\% at 
reaction threshold to 2.7\% at a beam energy of 
1.4~GeV. The background spectra determined from 
the analysis of the empty-target samples were 
then subtracted from our full-target spectra.

\begin{figure*}
\includegraphics[width=12.cm,height=11.cm,bbllx=.5cm,bblly=.5cm,bburx=19.5cm,bbury=16.cm]{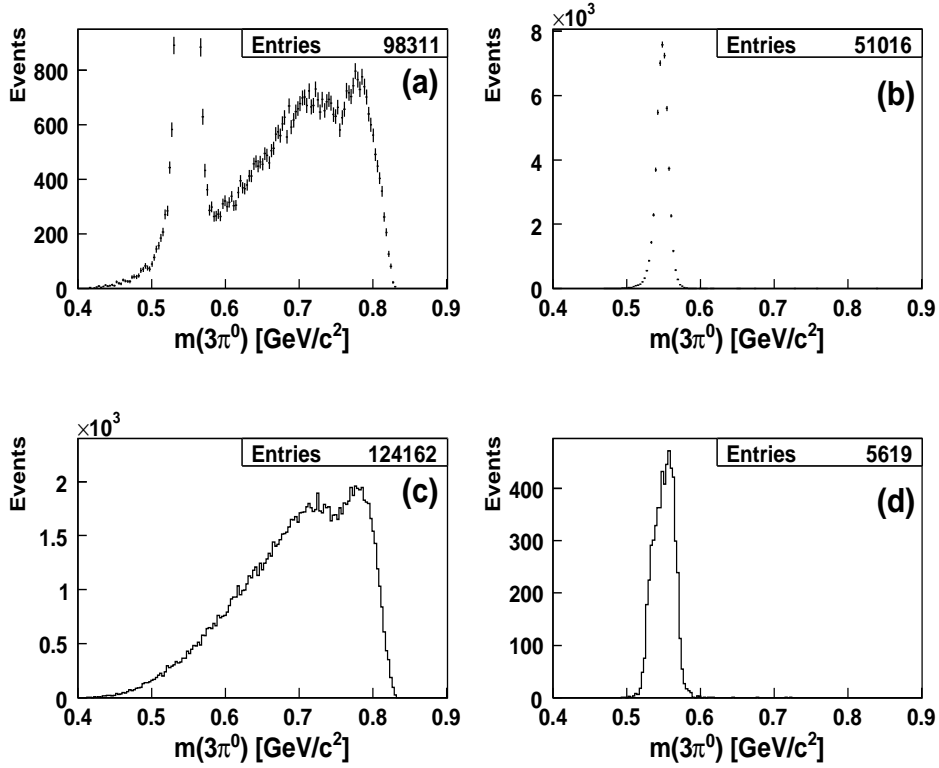}
\caption{Spectra of the $3\pi^0$ invariant mass 
	obtained from kinematic fitting to the 
	$\gamma p\to 3\pi^0p$ hypothesis for 
	the events with values of $E_\gamma$
	between 1150 and 1200~MeV:
	(a) experimental spectrum after the 
	subtraction of the random and 
	empty-target backgrounds,
	(b) experimental events selected also 
	as $\eta\to 3\pi^0$ candidates,
	(c) MC simulation for $\gamma p\to 
	3\pi^0p$,
	(d) events from the MC simulation for 
	$\gamma p\to 3\pi^0p$ selected also as 
	$\eta\to 3\pi^0$ candidates.}
	\label{fig:p3pi0_bg_bm}
\end{figure*}

Another source of background is the set of 
$\gamma p\to 3\pi^0p$ events that are not 
produced by $\eta\to 3\pi^0$ decays. When the 
invariant mass of the three neutral pions in 
the final state is sufficiently close to the 
mass of the $\eta$ meson, those events are 
selected as $\eta\to 3\pi^0$ candidates. A 
phase-space simulation of $\gamma p\to 3\pi^0p$ 
was used for the subtraction of this background. 
The fraction of this direct $3\pi^0$ background 
in our $\eta\to 3\pi^0$ candidates was 
determined in each photon-energy bin via the 
normalization of the MC simulation for $\gamma 
p\to 3\pi^0p$ to the corresponding experimental 
spectra (Fig.~\ref{fig:p3pi0_bg_bm}). The 
$3\pi^0$ invariant-mass distributions obtained 
from kinematic fitting to the $\gamma p\to 
3\pi^0p$ hypothesis were used for this purpose.
In Fig.~\ref{fig:p3pi0_bg_bm}, these 
distributions are shown for the experimental 
data and MC simulation at incident-photon 
energies between 1150~MeV and 1200~MeV.
%%%%%%%%%%%%%%%%%%%%%%%%%%%%%%%%%%%%%%%%%%%%%%%%%%%%%%
%%%%%%%%%%%%%%%%%%%%%%%%%%%%%%%%%%%%%%%%%%%%%
The experimental spectrum after the subtraction of 
random-coincidence and empty-target backgrounds is 
shown in Fig.~\ref{fig:p3pi0_bg_bm}(a). The 
experimental events that were then selected as 
$\eta\to 3\pi^0$ candidates are shown in 
Fig.~\ref{fig:p3pi0_bg_bm}(b).  The corresponding 
spectra obtained for the MC simulation of $\gamma 
p\to 3\pi^0p$ are shown in 
Figs.~\ref{fig:p3pi0_bg_bm}(c) and (d). The 
MC-simulation spectrum in 
Fig.~\ref{fig:p3pi0_bg_bm}(c) has a shape very 
similar to that of the experimental spectrum under 
the $\eta$ peak in Fig.~\ref{fig:p3pi0_bg_bm}(a), 
and its normalization to the data provides a 
good estimate of the direct $3\pi^0$ background 
in the $\eta\to 3\pi^0$ candidates. The fraction 
of direct $3\pi^0$ background that was 
subtracted from our experimental spectra was 
found to vary from 0.3\% at the $\eta$ threshold 
to 4.4\% at 1.4~GeV.

The large number of events accumulated allowed the
division of the data into 120 bins in $E_\gamma$.  
From the reaction threshold to an $E_\gamma$ of 
1008~MeV, the bin width was that of a single 
tagger channel ($\sim 4$~MeV). From 1008 to 
1238~MeV, two tagger channels were combined to a 
single energy bin. Above 1238~MeV, an energy bin 
included from three to eight tagger channels.
The $\gamma p\to\eta p$ differential cross 
sections were determined as a function of 
$\cos\theta$, where $\theta$ is the polar angle 
of the $\eta$ direction in the c.m. frame.  The 
$\cos\theta$ spectra at all energies were divided 
into 20 bins.

%%%%%%%%%%%%%%%%%%%%%%%%%%%%%%%%%%%%%%%%%%%%%%%%%%%%%%%%%
\begin{figure*}
\includegraphics[width=14.cm,height=5.5cm,bbllx=1.cm,bblly=0.5cm,bburx=19.cm,bbury=7.5cm]{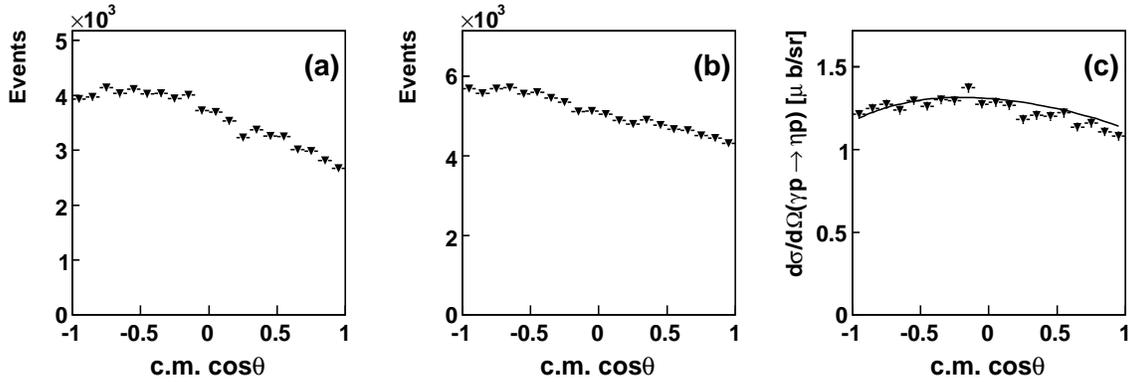}
\caption{$\cos\theta$ distributions in the c.m. frame 
    of $\eta$ photoproduction at $E_\gamma=760$~MeV ($W=1519$~MeV):
    (a) experimental $\gamma p\to\eta p\to 3\pi^0p$
    events;
    (b) MC simulation for the $\gamma p\to\eta p\to
    3\pi^0p$ events;
    (c) results for the $\gamma p\to\eta p$ 
	differential cross section compared to the 
	prediction by SAID (solution E429)
	\protect\cite{SAID} (solid line).
    }\label{fig:coseta3pi0_760mev}
\end{figure*}
%%%%%%%%%%%%%%%%%%%%%%%%%%%%%%%%%%%%%%%%%
\begin{figure*}
\includegraphics[width=14.cm,height=5.5cm,bbllx=1.cm,bblly=0.5cm,bburx=19.cm,bbury=7.5cm]{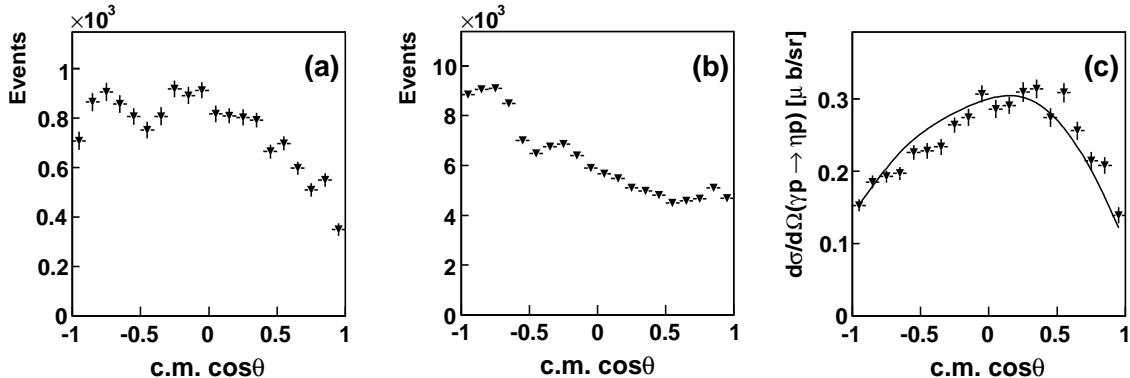}
\caption{Same as Fig.~\protect\ref{fig:coseta3pi0_760mev}
    but for $E_\gamma=1060$~MeV ($W=1694$~MeV).
    }\label{fig:coseta3pi0_1060mev}
\end{figure*}
%%%%%%%%%%%%%%%%%%%%%%%%%%%%%%%%%%%%%%%%%%%%%%

The determination of the differential cross sections
is illustrated in Figs.~\ref{fig:coseta3pi0_760mev}
and \ref{fig:coseta3pi0_1060mev} for incident-photon 
energies of 760 and 1060~MeV, respectively. The 
experimental distributions after subtraction of all 
background contributions are shown in 
Figs.~\ref{fig:coseta3pi0_760mev}(a) and
\ref{fig:coseta3pi0_1060mev}(a). The corresponding
distributions from the MC simulation are shown in
Figs.~\ref{fig:coseta3pi0_760mev}(b) and
\ref{fig:coseta3pi0_1060mev}(b). Since the 
simulated angular distribution is isotropic, the MC 
spectra reflect the experimental acceptance as a 
function of $\cos\theta$. The change in the 
acceptance with the incident-photon energy is 
caused by a more pronounced Lorentz boosting to 
forward angles, so that a larger fraction of 
particles impinge upon the overlap region between 
the CB and TAPS, where the metal framework of the 
CB aperture reduces detection efficiency.
To obtain the $\gamma p\to\eta p$ differential 
cross sections, the experimental distributions
were normalized for the acceptance, 
the $\eta\to 3\pi^0$ branching ratio, the photon 
beam flux, and the number of target protons. 
These differential cross sections are shown in 
Figs.~\ref{fig:coseta3pi0_760mev}(c) and 
\ref{fig:coseta3pi0_1060mev}(c).
The results are 
very close to the predictions of SAID~\cite{SAID} 
for the $\gamma p\to\eta p$ differential cross 
sections at the given energies.  These predictions 
are shown by the solid lines in the same figures.

%%%%%%%%%%%%%%%%%%%%%%%%%%%%%%%%%%%%%%%%%%%%
\section{Systematic Uncertainties}
\label{sec:Errs}

The results presented for the total, $\sigma_t(\gamma 
p\to\eta p)$, and differential cross sections include 
only statistical uncertainties. The largest 
contributions to the systematic uncertainty come from 
the calculation of the experimental acceptance by the 
MC simulation and from the determination of the 
photon-beam flux. A good test of the MC simulation is 
the determination of the $\gamma p\to\eta p$ 
differential cross sections using the two different 
modes of $\eta$ decays: $\eta\to 3\pi^0$ and 
$\eta\to\gamma\gamma$. As discussed above, the data 
used in the present analysis were taken with the 
trigger suppressing the events with low cluster 
multiplicity in the final state. To perform our test, 
we used a data sample that contained much fewer 
events, taken at a later stage with an almost open 
trigger. The results obtained for the 
$\gamma p\to\eta p$ total cross sections from the two
different decay modes of $\eta$ are in good
agreement within their statistical uncertainties,
the magnitude of which are $\sim 2$\% for $\eta\to 
3\pi^0$ and $\sim 1$\% for $\eta\to\gamma\gamma$.
They also agree with the high-precision $\gamma 
p\to\eta p$ results presented here. Based on the 
comparison of our own results with each other and 
with the existing data in the region of the 
$N(1535)1/2^-$ (the most well-known region), the 
general systematic uncertainty in our $\gamma 
p\to\eta p$ cross sections was estimated to be 4\%. 
To take into account the statistical uncertainties 
in the estimation of the tagging efficiency of every 
individual tagger channel, used for the 
photon-flux calculation, those uncertainties were 
added in quadrature with our general systematic 
uncertainty. The typical magnitudes of the 
statistical uncertainties in the tagging efficiencies
of the tagger channels for our data are between 
1.4\% at the $\eta$ threshold and 2.5\% at the 
largest energies. For every post-bremsstrahlung electron
detected by a tagger channel, the tagging efficiency
reflects a probability of the corresponding bremsstrahlung photon
to pass through the photon collimator and to reach the
target. The typical tagging efficiency of the tagger channels
in the present experiment varied between $67\%-71\%$. 
%%%%%%%%%%%%%%%%%%%%%%%%%%%%%%%%%%%%%%%%%%%%%%%%%%%%%
\begin{figure*}[th]
\centerline{
\includegraphics[height=0.48\textwidth, angle=90]{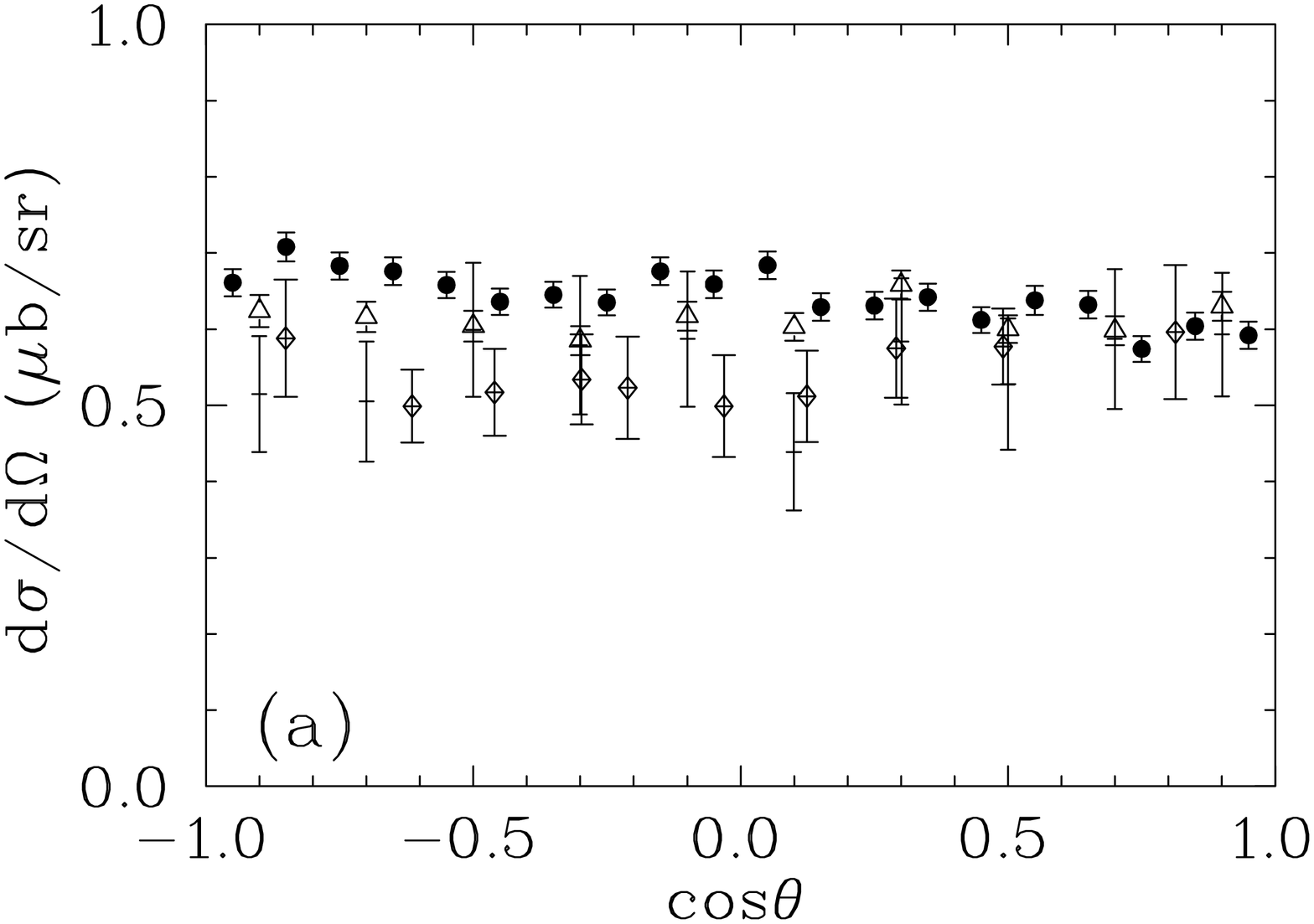}\hfill
\includegraphics[height=0.48\textwidth, angle=90]{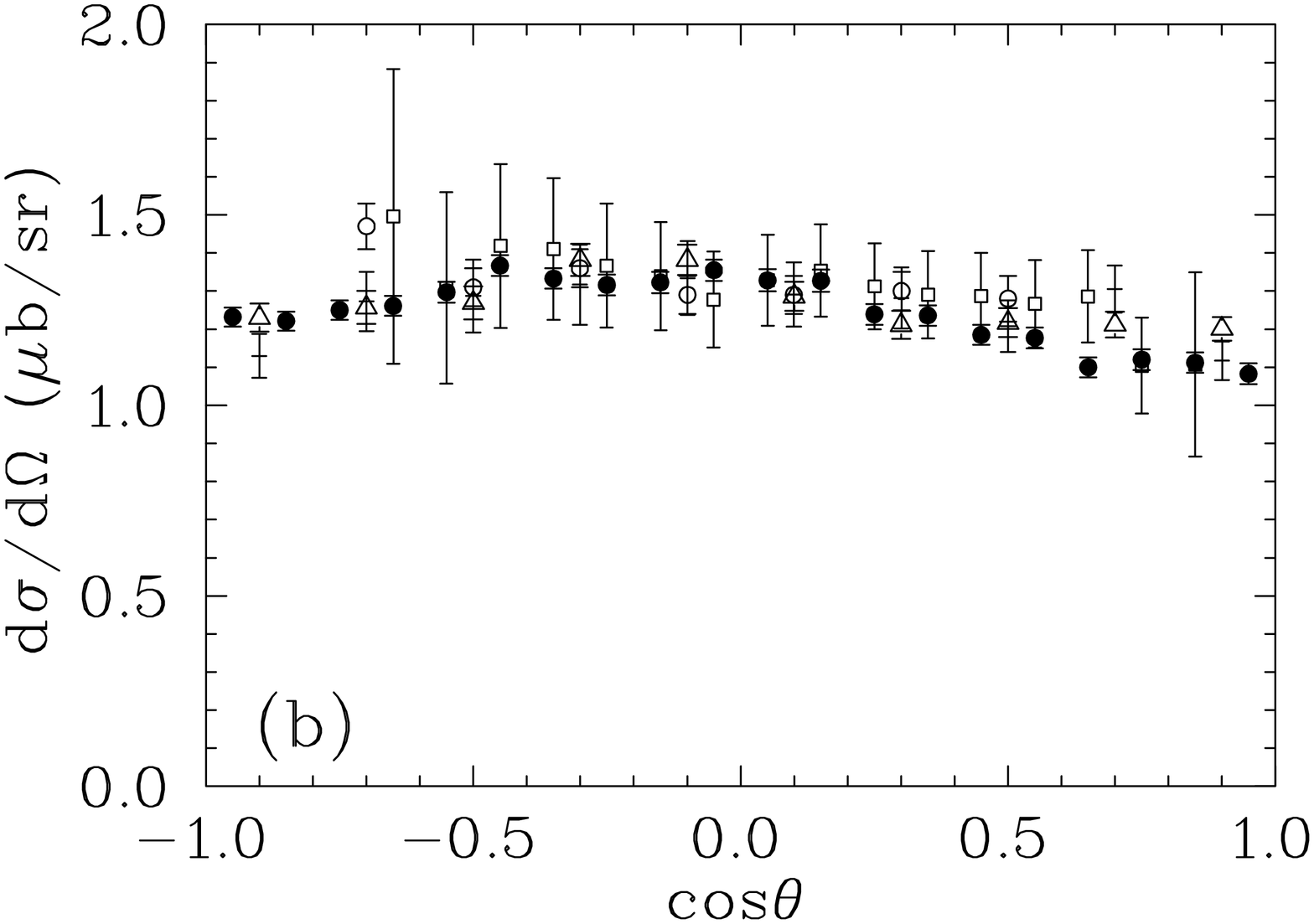}}
\centerline{
\includegraphics[height=0.48\textwidth, angle=90]{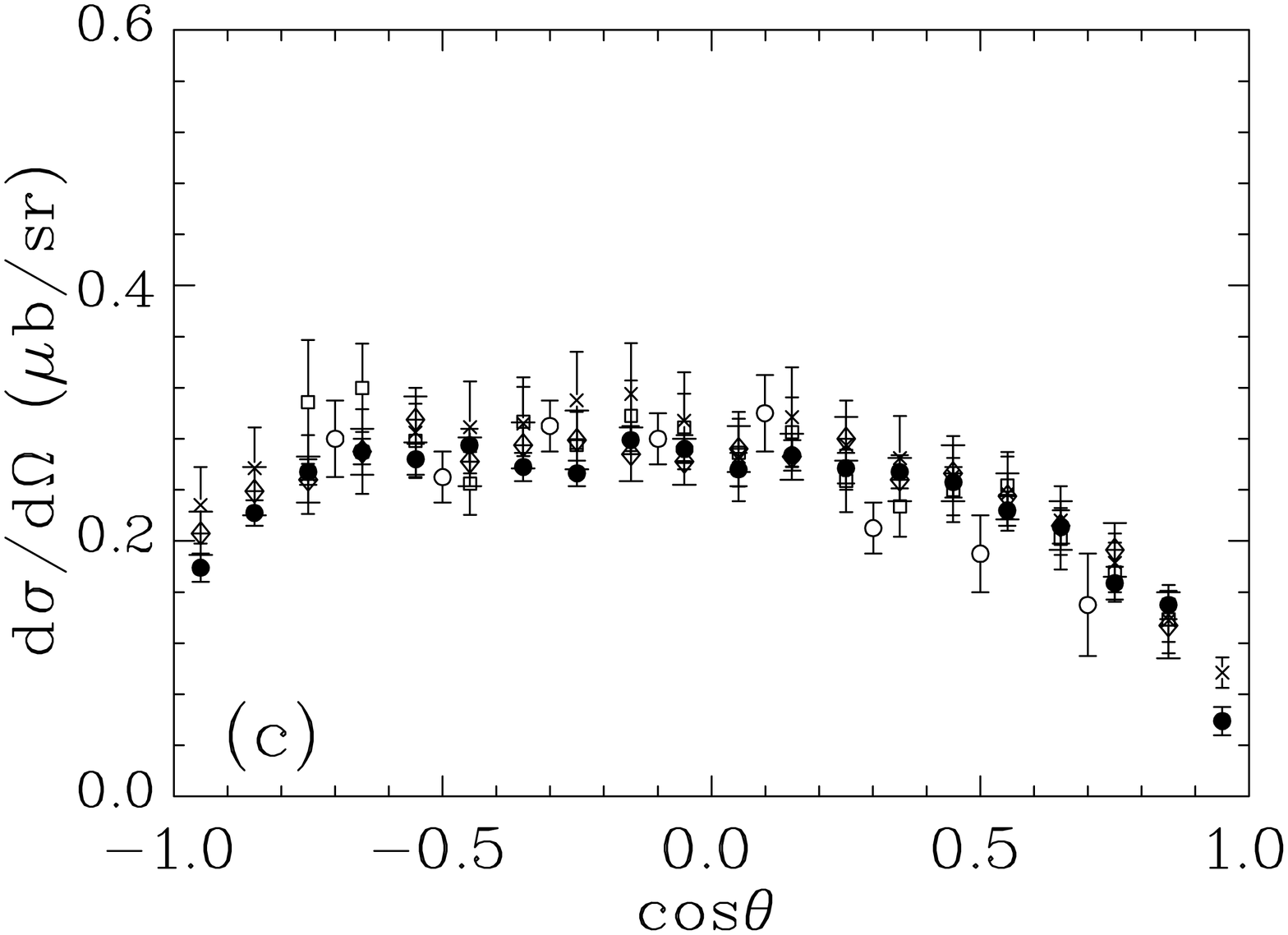}\hfill
\includegraphics[height=0.48\textwidth, angle=90]{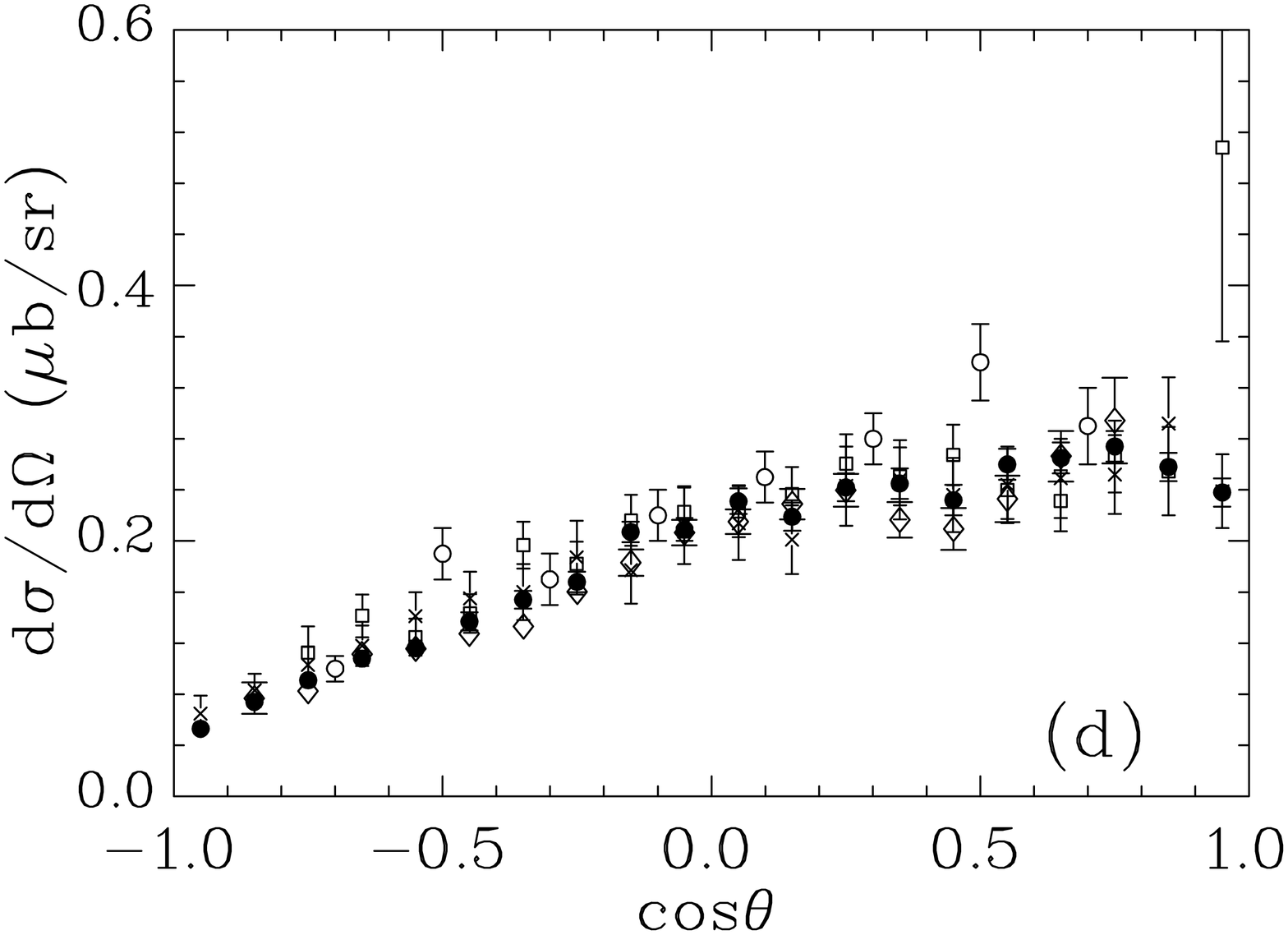}}
\caption{Differential cross sections for $\gamma p\to\eta p$
	as a function of $\cos\theta$, where $\theta$
	is the production angle of $\eta$ in the c.m. frame.  
	The present data (solid circles) are shown for four 
	energy bins:
	(a) $E_\gamma = 714.5\pm 2.1$~MeV, 
	(b) $772.9\pm 2.1$~MeV,
	(c) $1026.8\pm 3.7$~MeV, and
	(d) $1376.2\pm 9.7$~MeV.
	Previous data are shown for experiments at
	MAMI-B~\protect\cite{kr95} for
	$715.9\pm 5$~MeV and $775.3\pm 5$~MeV (open triangles); 
	CLAS-g1c~\protect\cite{du02} for $775\pm 25$~MeV,
	$1025\pm 25$~MeV, and $1375\pm 25$~MeV (open circles);
	CLAS-g11a~\protect\cite{wi09} for $1384\pm 10$~MeV 
	(open diamonds);
	GRAAL~\protect\cite{re02} for $714\pm 9$~MeV and 
	$1024\pm 9$~MeV (open diamonds with crosses);
	LNS~\protect\cite{na06} for $718.0\pm 10$~MeV and 
	$768.8\pm 10$~MeV (horizontal bars);
	CB-ELSA~\protect\cite{cr05} for $774\pm 25$~MeV,
	$1025\pm 25$~MeV, and $1374\pm 25$~MeV (open squares); and 
	CB-ELSA/TAPS~\protect\cite{cr09} for $1025\pm 25$~MeV 
	and $1375\pm 25$~MeV (crosses).
	Plotted uncertainties are statistical only. 
	\label{fig:g1}}
\end{figure*}

%%%%%%%%%%%%%%%%%%%%%%%%%%%%%%%%%%%%%%%%%%%%%%%
\section{Results}
\label{sec:Results}

Since our results for the $\gamma p\to\eta p$ 
differential cross sections consist of 2400 
experimental points, they are not tabulated in 
this publication, but are available in the SAID 
database~\cite{SAID} along with their 
uncertainties and the energy binning. In this 
section, we compare our results to the world 
data set.

In Fig.~\ref{fig:g1}, our differential cross 
sections for four incident-photon energies are 
compared to previous measurements made at 
similar energies~\cite{kr95,du02,wi09,re02,na06,cr05,cr09}.  
Some of these measurements~\cite{wi09,cr09,su09} 
are quite recent, demonstrating the general 
desire of the resonance-physics community to 
obtain new $\gamma N\to\eta N$ data, which are 
needed for a better determination of the 
properties of the $N^\ast$ states. The lowest 
energy shown, $E_\gamma = 714.5$~MeV ($W = 
1490.3$~MeV), is close to the $\eta$-production 
threshold. The second energy, $E_\gamma = 
772.9$~MeV ($W = 1526.7$~MeV), is at the maximum 
of the total cross section. The third energy, 
$E_\gamma = 1026.8$~MeV ($W = 1675.4$~MeV), 
is at a local minimum of the total cross section.  
The last energy shown, $E_\gamma = 1376.2$~MeV 
($W = 1860.9$~MeV) is close to the maximum of our 
incident-photon energy range. As seen in 
Fig.~\ref{fig:g1}, all our results are in 
reasonable agreement with the previous 
measurements, but our statistical uncertainties 
are much smaller and the energy binning much 
finer. Larger discrepancies are observed between 
the data obtained close to the $\eta$-production 
threshold, but this can be explained by the 
difference in the energy binning of the data 
sets, bearing in mind the rapidly rising cross 
section close to threshold. 
%%%%%%%%%%%%%%%%%%%%%%%%%%%%%%%%%%%%%%%%%%%%%%%%%%%%%%
\begin{figure*}[th]
\includegraphics[height=0.8\textwidth, angle=90]{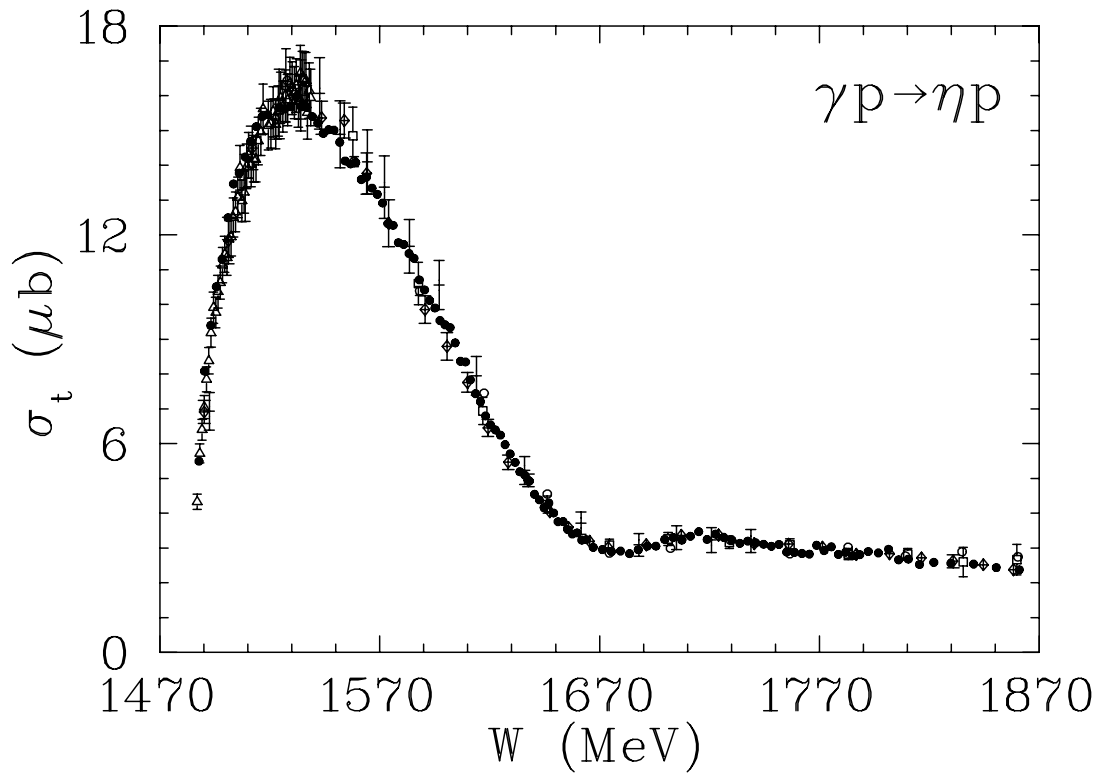}
\caption{Total cross section for $\gamma 
	p\to\eta p$ as a function of the c.m. 
	energy. Uncertainties plotted for our 
	data are statistical only.  Notation 
	of different data is the same as in 
	Fig.~\protect\ref{fig:g1}. \label{fig:g2}}
\end{figure*}
%%%%%%%%%%%%%%%%%%%%%%%%%%%%%%%%%%%%%%%%%%%%%%%%%%%%%%
\begin{figure*}[th]
\centerline{
\includegraphics[height=0.48\textwidth, angle=90]{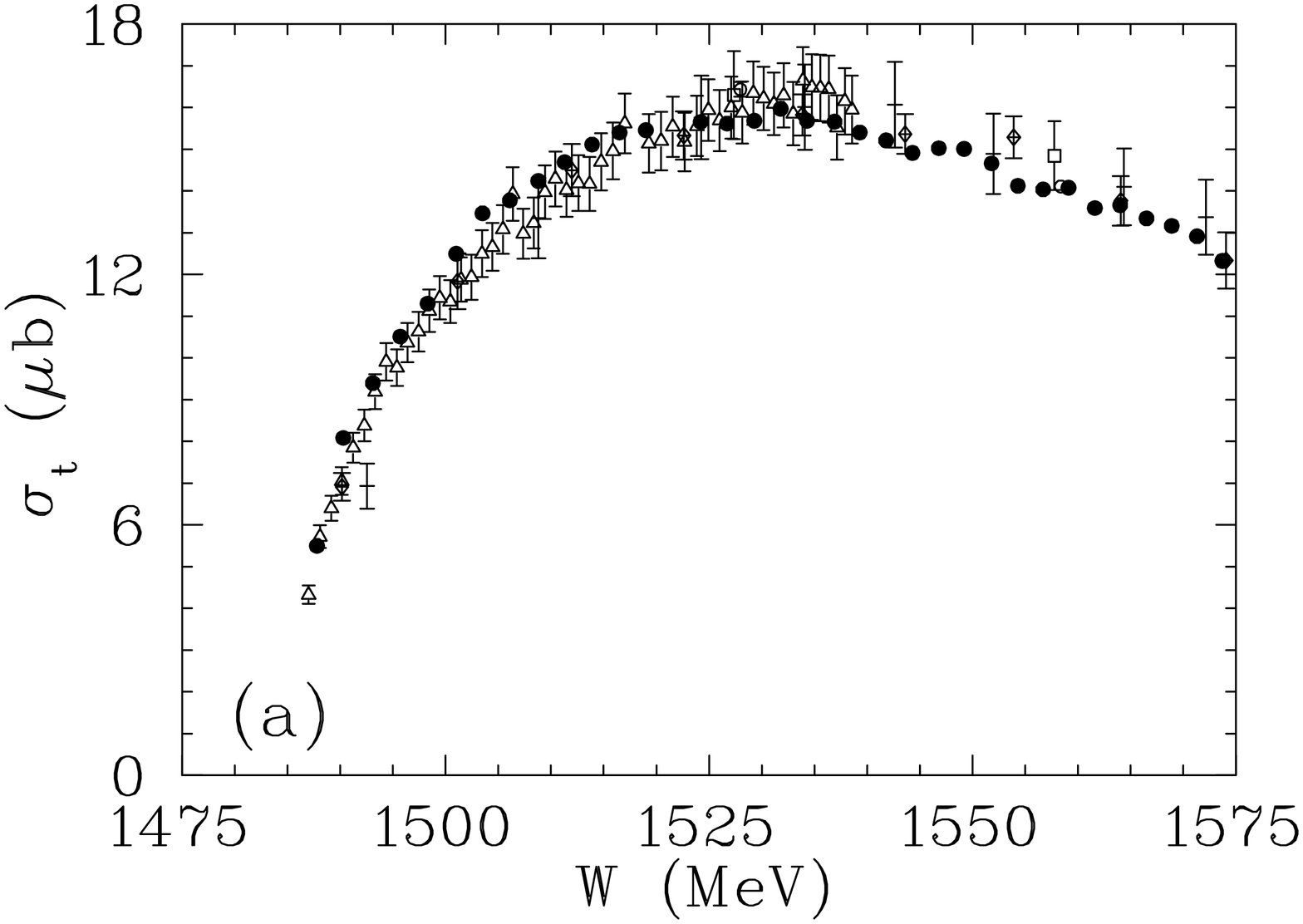}\hfill
\includegraphics[height=0.46\textwidth, angle=90]{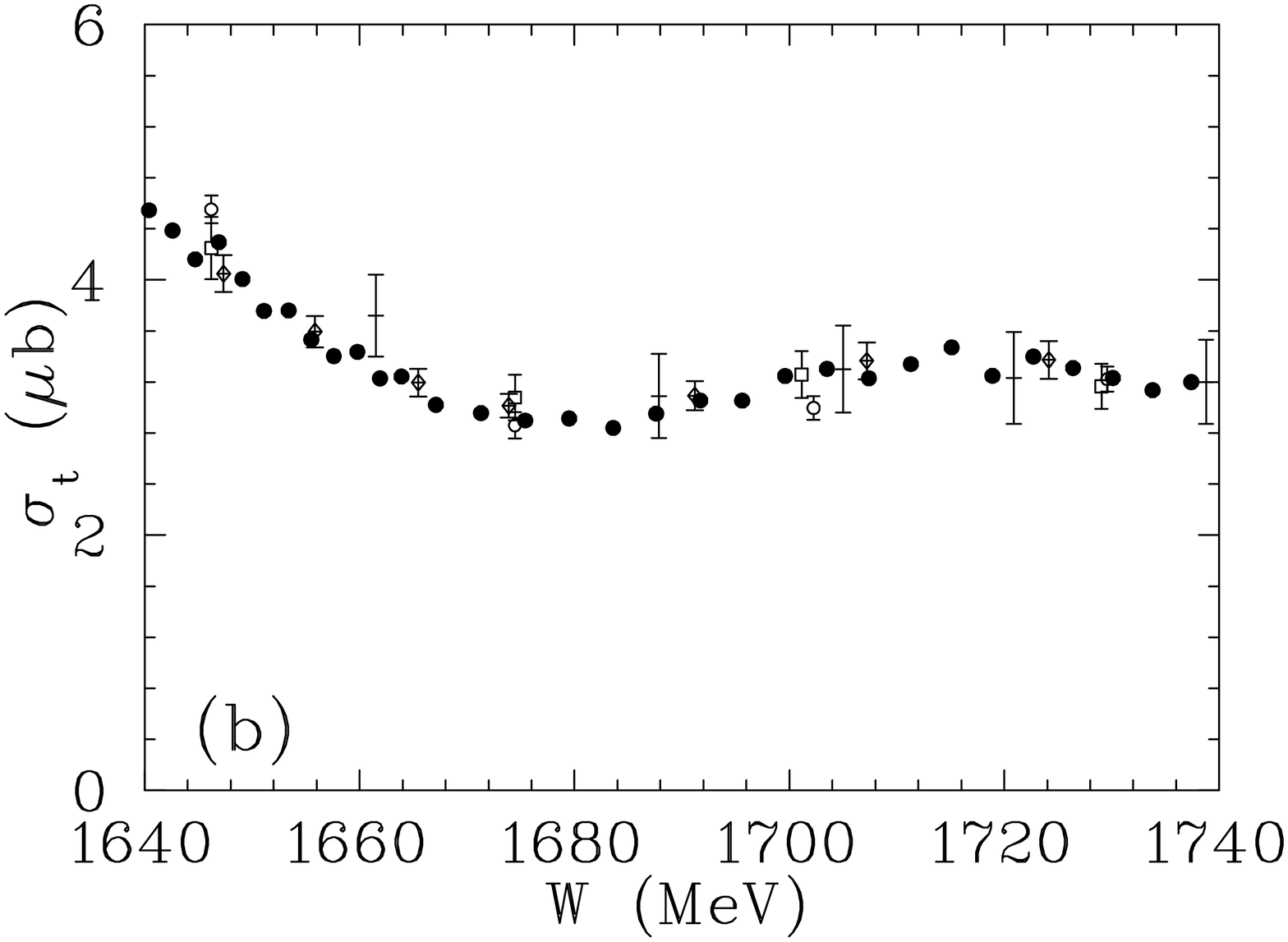}}
\caption{Same as Fig.~\protect\ref{fig:g2} but 
        for narrower $W$ ranges: (a) from the 
	threshold to the maximum of the total 
	cross section and (b) around a shallow dip,
	$W=1680$~MeV. \label{fig:g2a}}
\end{figure*}
%%%%%%%%%%%%%%%%%%%%%%%%%%%%%%%%%%%%%%%%%%%%%%%%%%%%%

The present total cross sections for $\gamma p\to\eta 
p$ are obtained by integration of the differential 
cross sections.  In Fig.~\ref{fig:g2}, our total 
cross sections are compared with previous 
measurements~\cite{kr95,du02,wi09,re02,na06,cr05,cr09} 
over the full energy range presently measured. A part 
of this distribution is repeated in Fig.~\ref{fig:g2a}(a), 
showing the range from the threshold to the 
$N(1535)1/2^-$ maximum in more detail.  Our results 
for the total cross sections are in general agreement 
with the major previous results.  
The energy range lying above a c.m. energy of 1640~MeV 
(see also Fig.~\ref{fig:g2a}(b)) is especially important 
for untangling the six overlapping $N^\ast$ states and 
investigation of a possible narrow $N^\ast$ state in the 
mass range $\sim 1680$~MeV. The $N^\ast(1680)$ was 
extracted in Ref.~\cite{p11} from the $\pi N$ PWA and 
suggested to be a member of the exotic anti-decuplet. 
Indeed a resonant bump at $\sim 1680$~MeV is observed in 
quasi-free $\gamma n\to\eta n$~\cite{graal,CB-Elsa,Lns}.  
However, for this reaction, the measured width of the 
bump was dominated by the experimental energy resolution. 
Inspection of our $\sigma_t(\gamma p\to\eta p)$ energy 
dependence shows no evidence for a narrow bump related 
to a $N^\ast(1680)$ state in $\eta$ photoproduction on a 
free proton.  Rather our data show the existence of a 
shallow dip near $W=1680$~MeV. However, such a situation 
may not contradict the existence of a narrow bump in 
$\eta$ photoproduction on a neutron, since $\gamma n$ 
and $\gamma p$ couplings of the $N^\ast(1680)$ can be 
essentially different (as for an anti-decuplet member).

\begin{figure}[th]
\includegraphics[height=0.48\textwidth, angle=90]{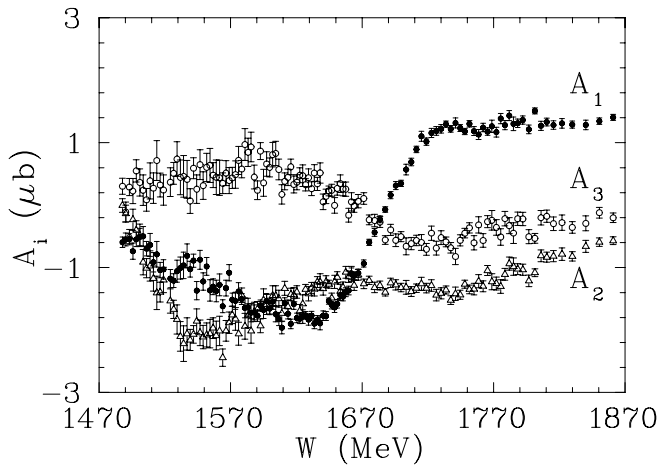}
\caption{Dominant Legendre coefficients from
	the fits to our differential cross sections. 
	The coefficients are plotted as a function 
	of the c.m. energy; $A_1$ is shown by solid 
	circles, $A_2$ by open triangles, and $A_3$ 
	by open circles. \label{fig:g2b}}
\end{figure}

The full angular coverage of our differential cross 
sections allied with the small statistical 
uncertainties allows a reliable determination of 
the Legendre coefficients $A_i$,which was difficult
to do with the previous data.
This unprecedented detail of the energy dependence of 
the Legendre coefficients will be indispensable in 
untangling the properties of the $N^\ast$ states 
lying in the present energy range.
In Fig.~\ref{fig:g2b}, 
we illustrate Legendre coefficients $A_1 - A_3$
(higher orders are relatively insignificant) as a 
function of the c.m. energy.
The swing in $A_1$ from negative to positive values
in the vicinity of $W=1680$~MeV is intriguing.
Since the first coefficient, $A_0$, simply reflects
the magnitude of the total cross section, it is not shown.

%%%%%%%%%%%%%%%%%%%%%%%%%%%%%%%%%%%%%%%%%%%%%%%%%%%%%%

%%%%%%%%%%%%%%%%%%%%%%%%%%%%%%%%%%%%%%%%%%%%%%%%%%%%%%
\section{Impact of the data on PWA}
\label{sec:Fit}

To gauge the influence of our data and
their compatibility with previous measurements,
our differential cross sections have been included
in a number of fits using the full SAID database
for $\gamma p\to\eta p$ up to $E_\gamma=2.9$~GeV.
The impact of our data on the SAID PWA can be
understood from the comparison of the new SAID
fit GE09, which involves our data, with the 
previous SAID fit E429~\cite{SAID}.  The other 
data included in the GE09 fit involve all 
previously published data except recent 
CLAS-g11a~\cite{wi09}, CB-ELSA/TAPS~\cite{cr09},
and LEPS~\cite{su09} differential cross sections. 
Our data were also included in the PWA under the 
Reggeized $\eta$-MAID model (Regge-MAID)
\cite{ReggeMAID} that was extended to a photon 
energy $E_\gamma=3.7$~GeV ($W=2.8$~GeV) by adding 
new resonances in the $s$-channel. Besides the 
resonances used in the original
Regge-MAID~\cite{ReggeMAID}, the new model 
includes five additional states from the fourth
resonance group, namely $N(1900)3/2^+$, 
$N(2000)5/2^+$, $N(2080)3/2^-$, $N(2090)1/2^-$, 
and $N(2100)1/2^+$, which are needed to describe 
the latest data from CLAS-g11a~\cite{wi09} and 
CB-ELSA/TAPS~\cite{cr09}. The influence of these
five states on the description of our data is 
very small. For the MAID solution without our 
data, we choose the $\eta$-MAID fit~\cite{etaMAID}, 
in which $E_\gamma$ is limited to 1.9~GeV 
($W<2.1$~GeV). The $\eta$-MAID analysis involves 
only the data published up to 2002.  The details 
of the new SAID and Regge-MAID PWAs will be the 
subject of future publications.

To search for the minimum $\chi^2$ value in the SAID 
fits, an overall rescaling of the differential cross 
sections was permitted within limits specified by 
the experimental systematic uncertainties~\cite{sm02}.  
A similar rescaling of the data, but without possible 
adjustment of the partial waves, was applied in the 
$\eta$-MAID and Regge-MAID fits. Comparison of the 
$\chi^2$ values from the two SAID fits, E429 and GE09, 
and from the two MAID fits, $\eta$-MAID and new 
Regge-MAID, is given in Table~\ref{tab:tbl1}.
%%%%%%%%%%%%%%%%%%%%%%%%%%%%%%%%%%%%%%%%%%%%%%%%%%%%%%%%%%%
\begin{table}[th]
\caption{Comparison of the $\chi^2$ values from the 
    fits to the $\eta$-photoproduction data for  
    the two SAID solutions, E429 and GE09, and
    the two MAID solutions, $\eta$-MAID and new
    Regge-MAID.  The fits were performed for 
    different $E_\gamma$ ranges and amount of data 
    in the databases used to fulfill the fits. See 
    text for more details. \label{tab:tbl1}}
\vspace{2mm}
\begin{tabular}{|c|c|c|c|}
\colrule
Solution     &$E_\gamma$ (MeV)&Data& $\chi^2$/Data \\
\colrule
GE09         & $<2900$        & 4211 & 1.5 \\
E429         & $<2900$        & 1811 & 1.6 \\
$\eta$-MAID  & $<1900$        & 1329 & 3.1 \\
Regge-MAID   & $<3700$        & 4161 & 5.9 \\
\colrule
\end{tabular}
\end{table}
%%%%%%%%%%%%%%%%%%%%%%%%%%%%%%%%%%%%%%%%%%%%%%%%%%%%%
The separate contributions of individual data sets
to the total $\chi^2$ value are listed for each of 
the four $\gamma p\to\eta p$ analyses in 
Table~\ref{tab:tbl2}.
%%%%%%%%%%%%%%%%%%%%%%%%%%%%%%%%%%%%%%%%%%%%%%%%%%%
\begin{table*}[th]
\caption{Individual contributions of different 
	measurements of the differential cross sections
	to the total $\chi^2$ value for the E429, GE09, 
    $\eta$-MAID, and new Regge-MAID analyses of 
    $\gamma p\to\eta p$ data. See text for details. 
    \label{tab:tbl2}}
\vspace{2mm}
\begin{tabular}{|c|c|c|c|c|c|c|c|}
\colrule
Experiment  & $E_\gamma$&  $W$       & $\theta$ &Stat/Syst& Data & $\chi^2$/data             & Ref \\
            & (MeV)     & (MeV)      &  (deg)   &    (\%) &      & E429$-$GE09$-$$\eta$-MAID$-$Regge-MAID & \\
\colrule
CB-MAMI-C   & 707$-$1402& 1487$-$1875&  0$-$180 &    2/4  & 2400 &5.5$-$1.4$-$6.8$-$3.2 & This Work\\
GRAAL       & 714$-$1477& 1491$-$1913& 32$-$162 &   10/3  &  487 &1.2$-$0.7$-$2.1$-$1.2 & \protect\cite{re02}\\
TAPS-MAMI-B & 716$-$ 790& 1493$-$1539& 26$-$154 &    4/4  &  100 &1.4$-$1.4$-$1.2$-$4.9 & \protect\cite{kr95}\\
LNS         & 718$-$1142& 1494$-$1740& 26$-$154 &    2/6  &  180 &0.9$-$1.5$-$0.9$-$2.3 & \protect\cite{na06}\\
CB-ELSA     & 774$-$2887& 1530$-$2511& 18$-$139 &    2/15 &  631 &1.3$-$1.3$-$3.1$-$2.6 & \protect\cite{cr05}\\
CLAS-g1c    & 775$-$1925& 1530$-$2121& 46$-$134 &    3/5  &  190 &2.3$-$2.3$-$5.6$-$5.4 & \protect\cite{du02}\\
CB-ELSA/TAPS& 875$-$2522& 1590$-$2372& 18$-$162 &    4/10 &  680 &2.6$-$2.6$-$11.2$-$3.3& \protect\cite{cr09}\\
CLAS-g11a   &1044$-$2861& 1690$-$2502& 33$-$148 &    7/11 &  979 &4.3$-$5.7$-$12.0$-$9.5& \protect\cite{wi09}\\
%LEPS        &1600$-$2400& 1972$-$2322&129$-$180 &    9/12 &   32 &34 $-$51 $-$1503& \protect\cite{su09}\\
\colrule
\end{tabular}
\end{table*}
%%%%%%%%%%%%%%%%%%%%%%%%%%%%%%%%%%%%%%%%%%%%%%%%%%%%%%%%%
These indicate that the more recent data sets 
display a greater degree of consistency.
However, the description of the CLAS-g11a data is 
worse with the new fit GE09, compared to the previous 
solution E429. Although in the overlapping energy 
range $W=1690$ to 1875~MeV, our data and the 
CLAS-g11a data are in good agreement.

\begin{figure*}[th]
\includegraphics[height=0.8\textwidth, angle=90]{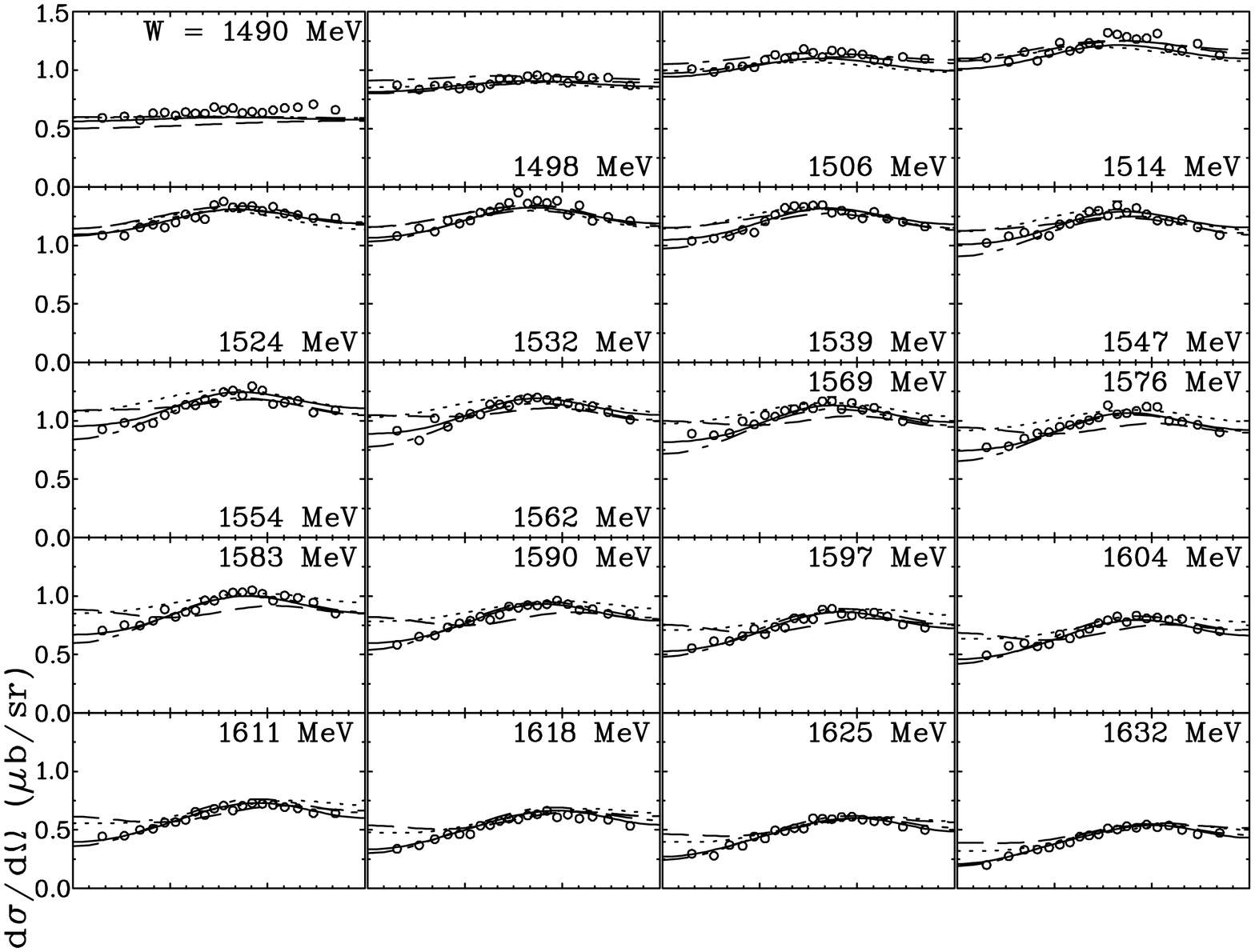}\\%\hfill
\includegraphics[height=0.8\textwidth, angle=90]{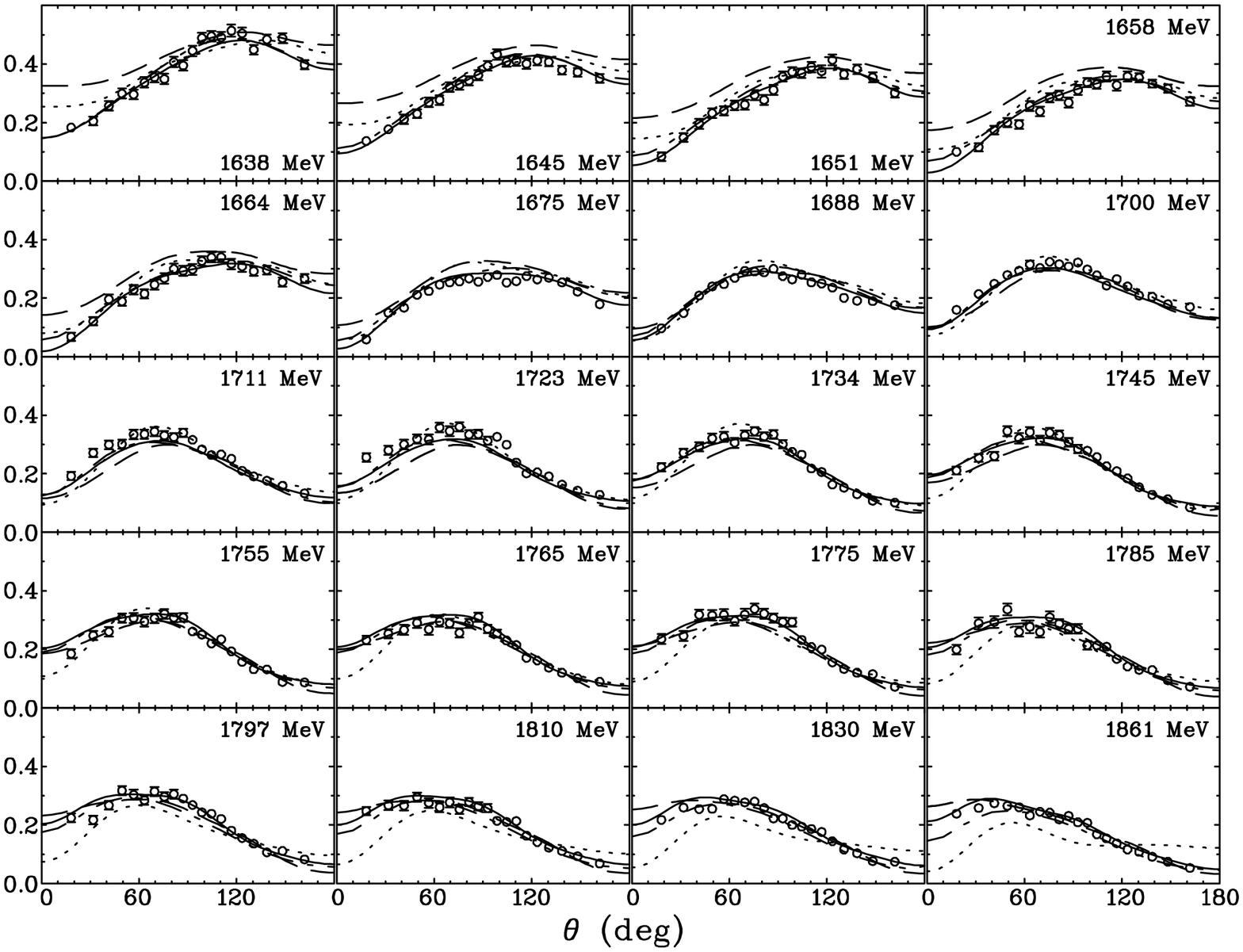}
\caption{Present differential cross sections 
    (open circles) for $\gamma p\to\eta p$ as a 
    function of $\theta$, the $\eta$ production 
    angle in the c.m. frame, over the range of 
    c.m. energies $W$ presently measured. The 
    plotted uncertainties are statistical only.
    The curves denote SAID solution GE09 (solid 
    line), SAID solution E429 (dashed line), 
    $\eta$-MAID solution (dotted line), and 
    Regge-MAID solution (dot-dashed line).
    \label{fig:g3}}
\end{figure*}

\begin{figure*}[th]
\centerline{
\includegraphics[height=0.3\textwidth, angle=90]{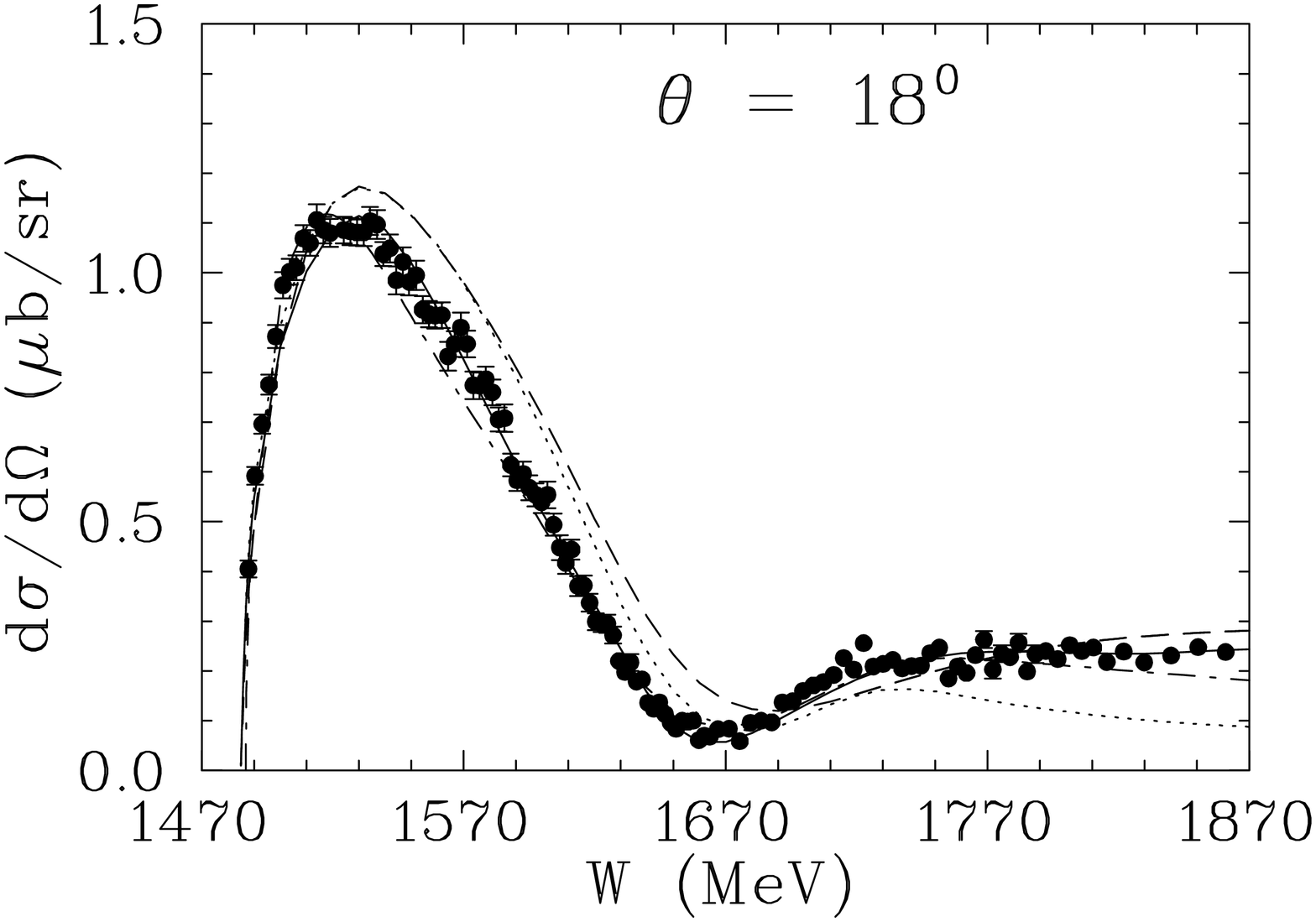}\hfill
\includegraphics[height=0.3\textwidth, angle=90]{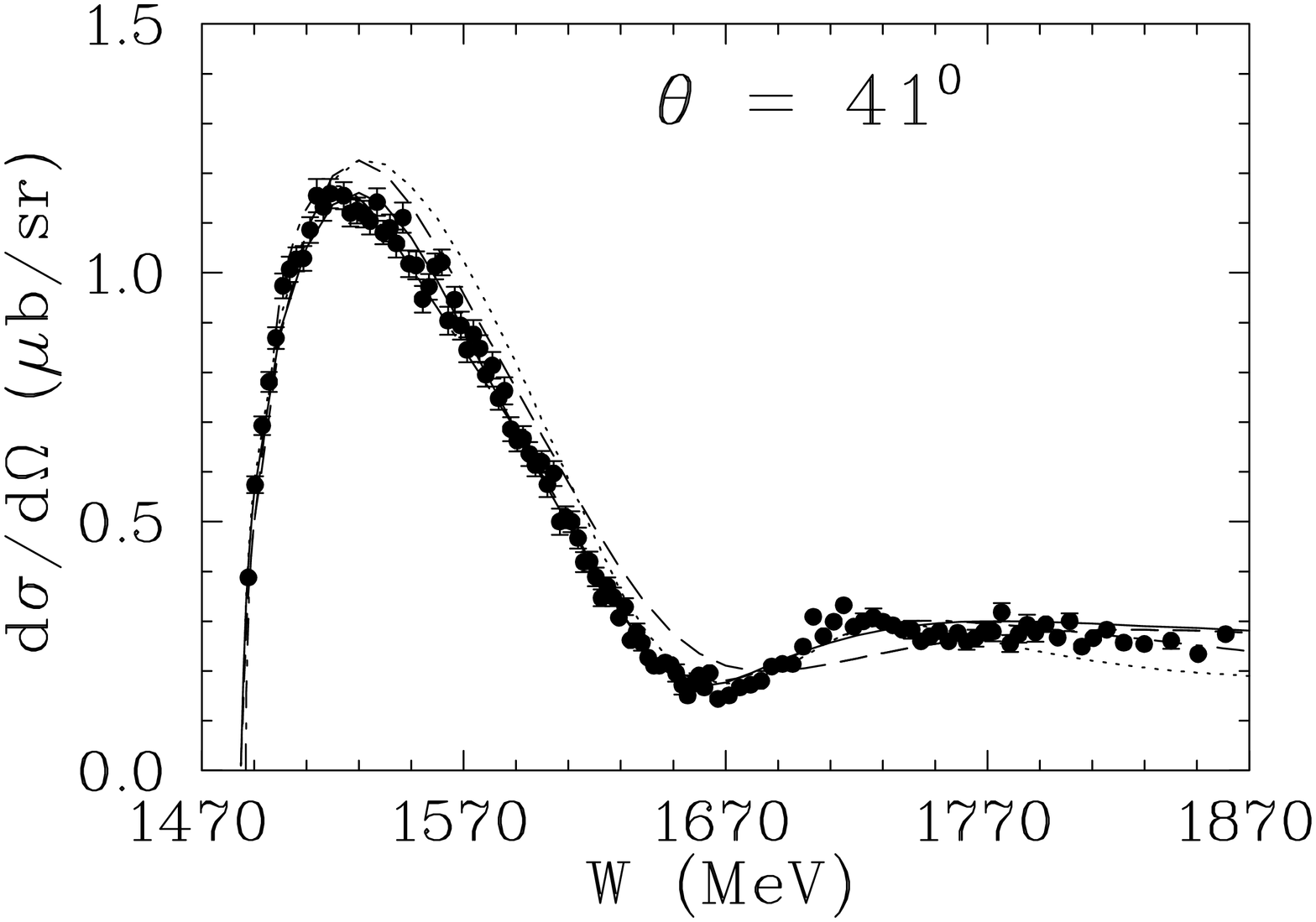}\hfill
\includegraphics[height=0.3\textwidth, angle=90]{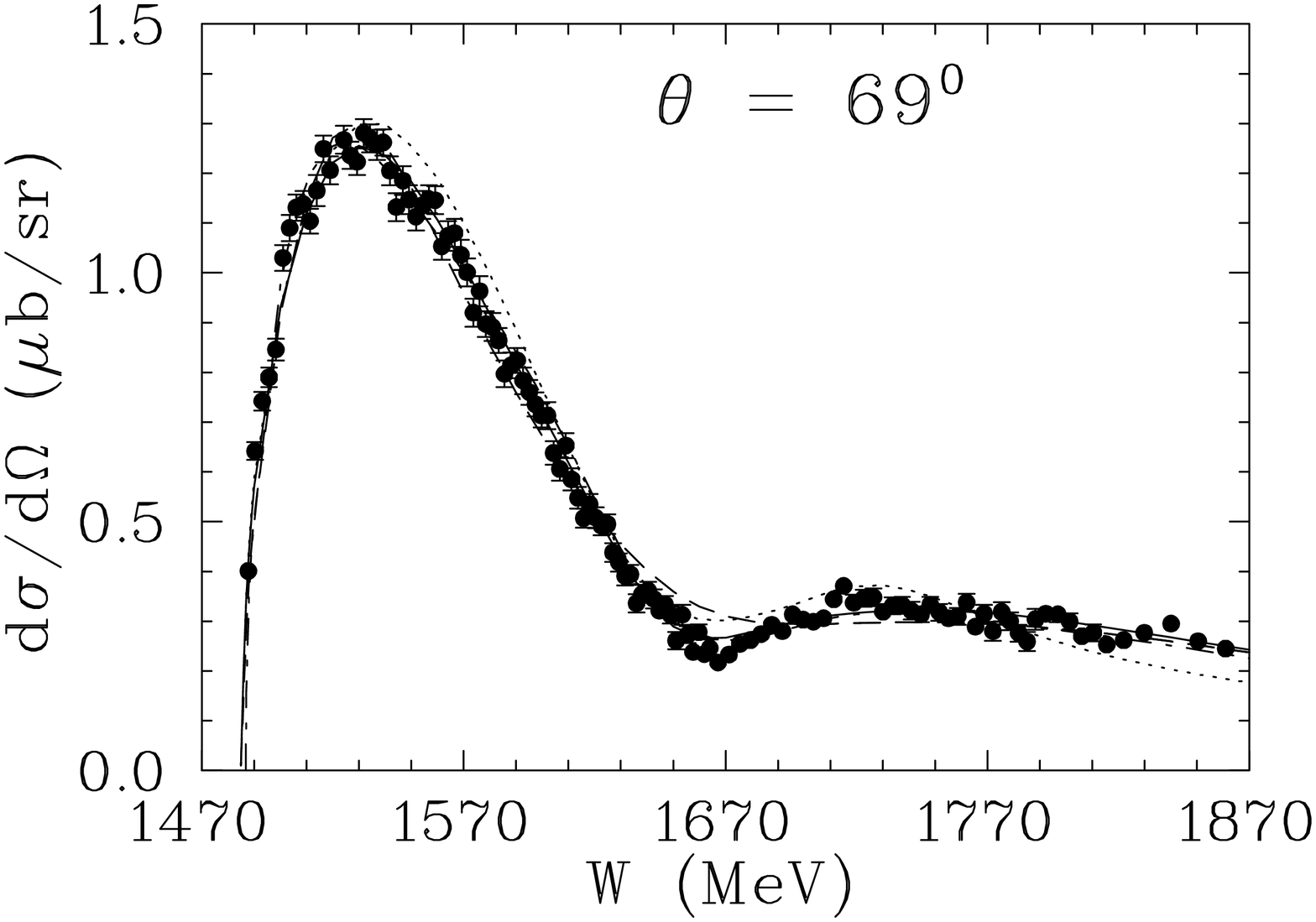}}
\centerline{
\includegraphics[height=0.3\textwidth, angle=90]{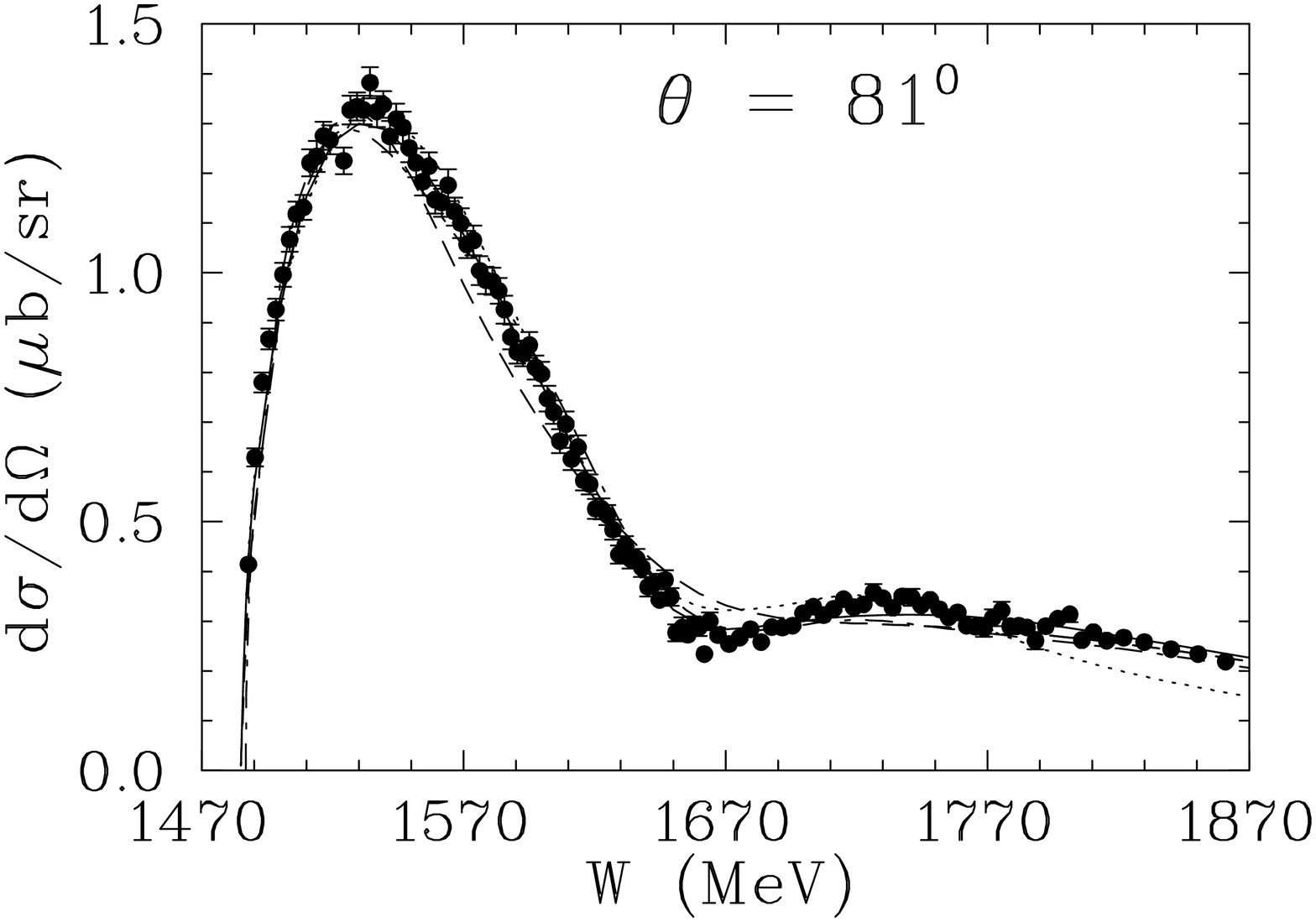}\hfill
\includegraphics[height=0.3\textwidth, angle=90]{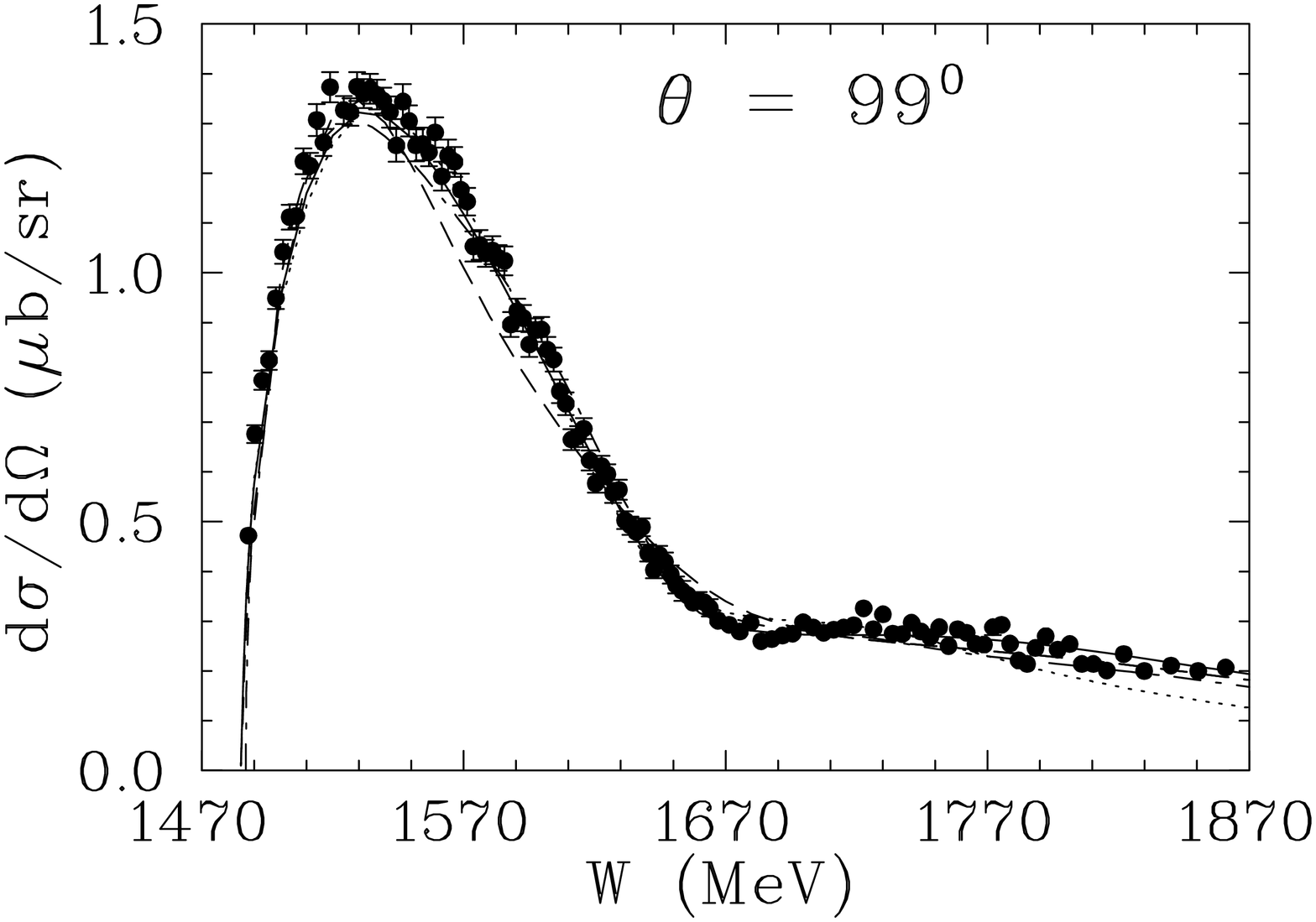}\hfill
\includegraphics[height=0.3\textwidth, angle=90]{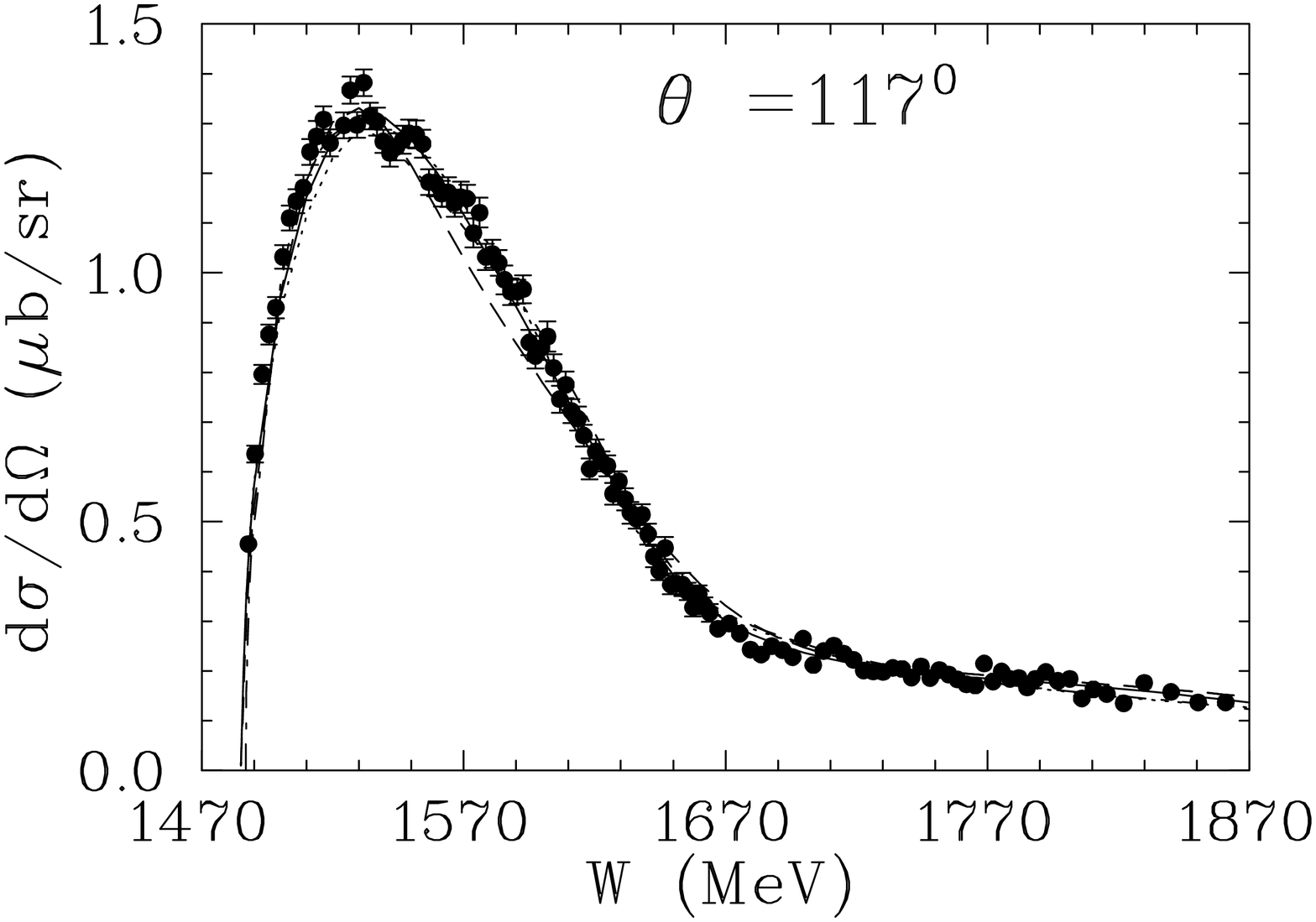}}
\centerline{
\includegraphics[height=0.3\textwidth, angle=90]{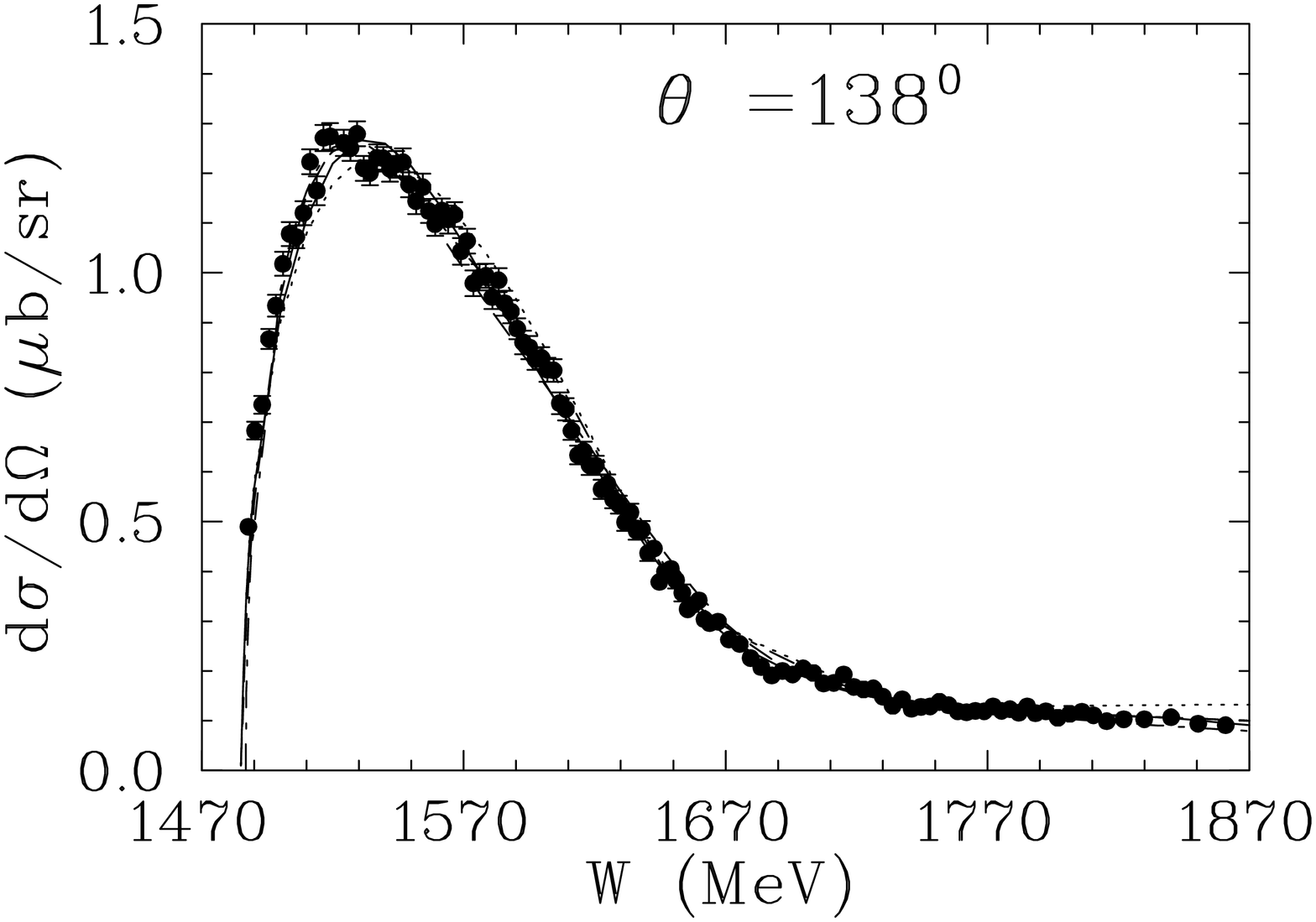}\hfill
\includegraphics[height=0.3\textwidth, angle=90]{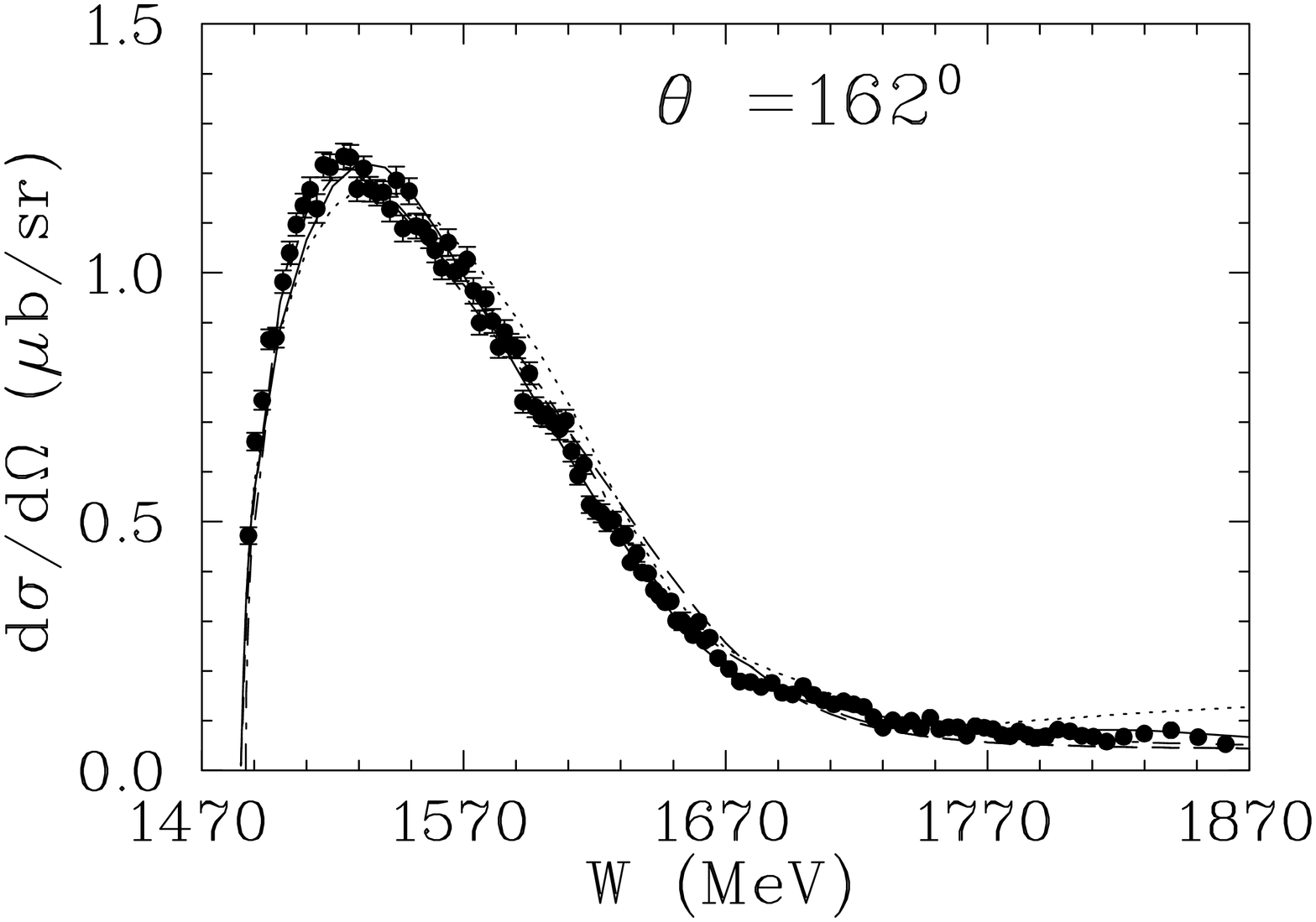}\hfill
\includegraphics[height=0.3\textwidth, angle=90]{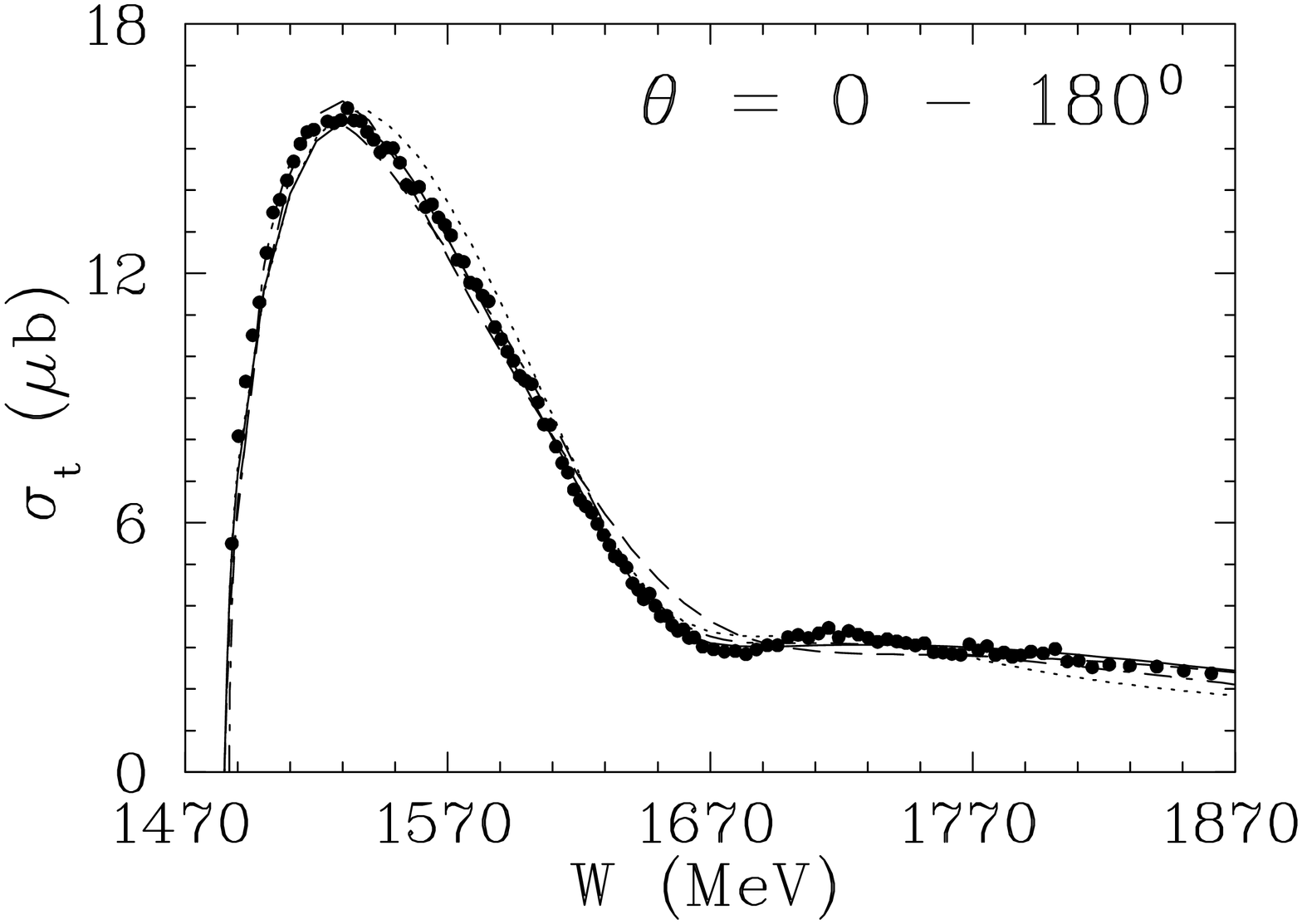}}
\caption{Fixed-angle excitation functions 
    for $\gamma p\to\eta p$ as a function 
    of the c.m. energy $W$ shown for eight 
    values of the $\eta$ production angle 
    and for the full angular range. Our 
    data are shown by solid circles. The 
    plotted uncertainties are statistical 
    only. The notation of the PWA solutions 
    is the same as in 
    Fig.~\protect\ref{fig:g3}. \label{fig:g4}}
\end{figure*}

In Fig.~\ref{fig:g3}, we show our differential cross 
sections at 40 energies and compare them with the results 
of each of the four PWA fits. In Fig.~\ref{fig:g4}, a 
similar comparison 
is made for the excitation functions for eight production 
angles and for the full angular range. The number of the 
distributions shown is enough to illustrate the quality of 
our data, the main features of the $\gamma p\to\eta p$ 
dynamics at the measured energy range, and the impact of 
the present data on PWAs.
The most noticeable effect of the present data on the new 
GE09 and Regge-MAID is due to very good measurements of the 
forward-angle cross sections for W in the range between 
1545 and 1675~MeV. Earlier, this forward region either 
had been measured with worse accuracy or could only be 
reached by extrapolation.

\begin{figure*}[th]
\centerline{
\includegraphics[height=0.38\textwidth, angle=90]{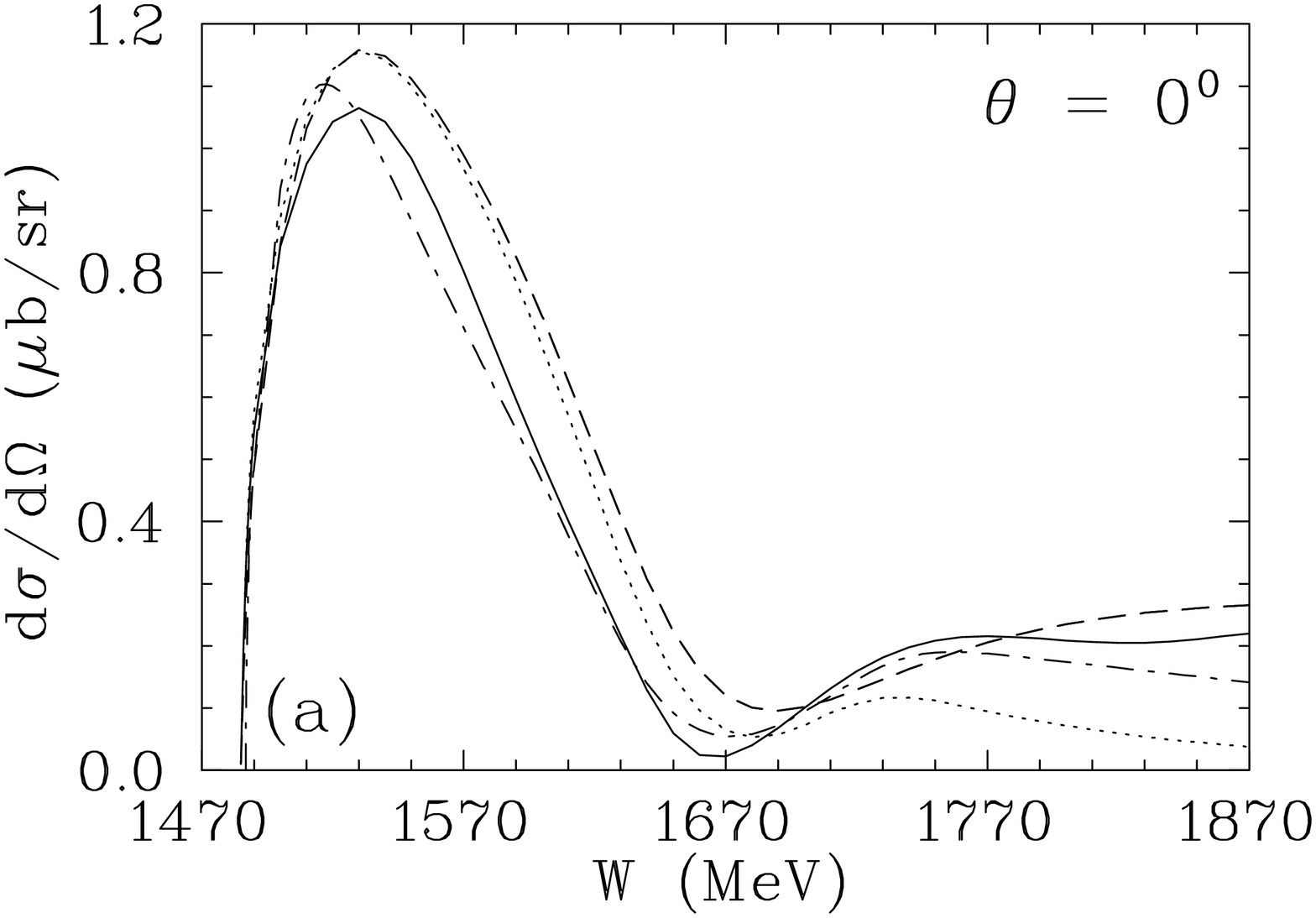}\hfill
\includegraphics[height=0.38\textwidth, angle=90]{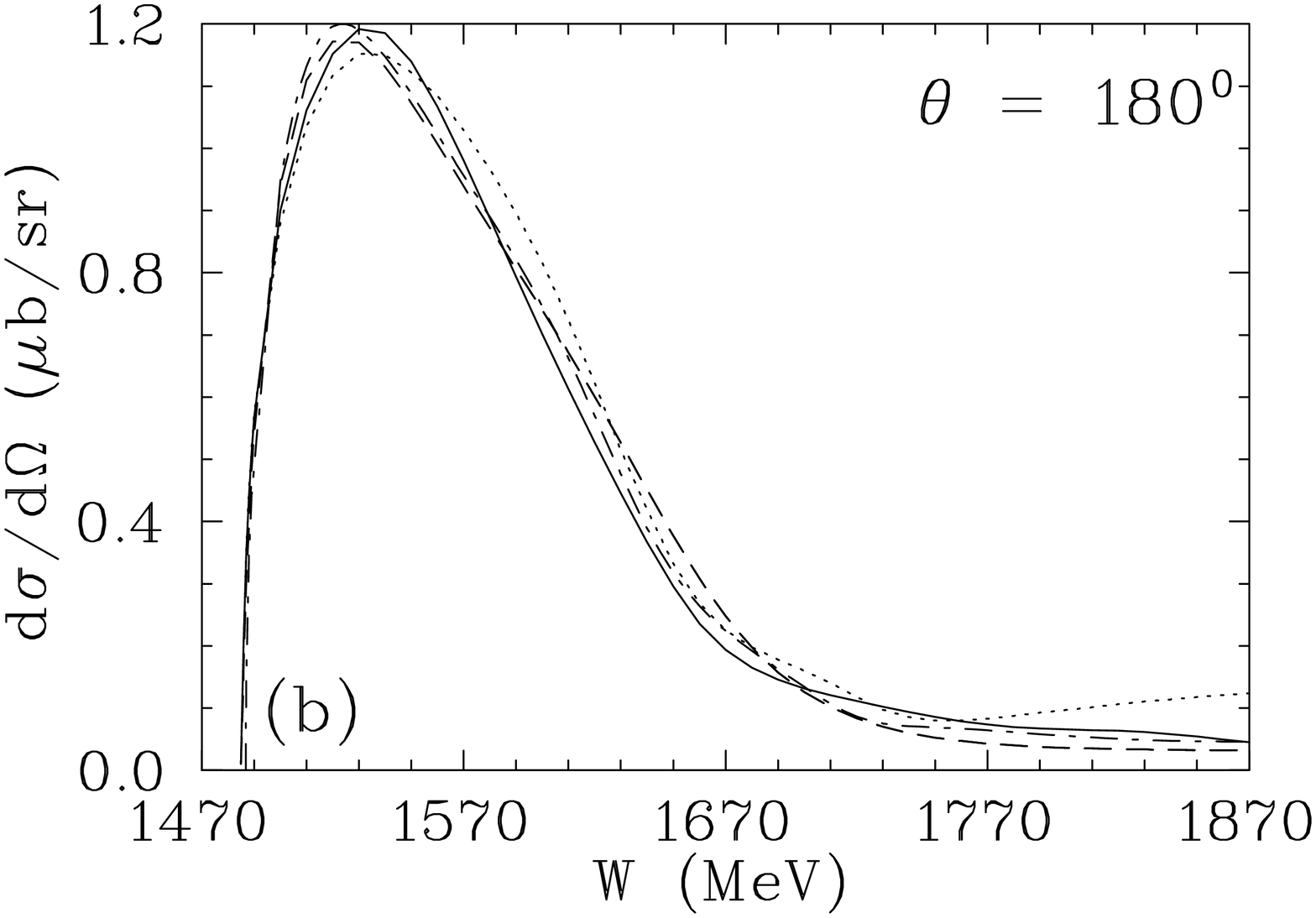}}
\caption{PWA predictions for the $\gamma p\to\eta p$ 
	excitation function at the extreme (forward 
	and backward) production angles of $\eta$, 
	shown as a function of the c.m. energy $W$. 
	Notation of the PWA predictions is the same 
	as in Fig.~\protect\ref{fig:g3}. 
	\label{fig:g5}}
\end{figure*}

For completeness, in Fig.~\ref{fig:g5} we compare the GE09, 
E429, $\eta$-MAID, and new Regge-MAID solutions for the 
$\gamma p\to\eta p$ excitation function at the extreme 
production angles: forward ($\theta = 0^\circ$) and 
backward ($\theta = 180^\circ$). The new data along 
with the new fits definitely indicate the existence of a 
dip structure around $W=1670$~MeV, which has already been 
seen in our total cross section (see Fig.~\ref{fig:g2}) 
and becomes very pronounced at forward production angles 
of $\eta$ (see Figs.~\ref{fig:g4} and \ref{fig:g5}). This 
feature was missed or questionable in the analysis of the 
previous data.
%%%%%%%%%%%%%%%%%%%%%%%%%%%%%%%%%%%%%%%%%%%%%%%%%%%%%%%%%%
%%%%%%%%%%%%%%%%%%%%%%%%%%%%%%%%%%%%%%%%%%%%%%%%%%%%%%%%%%
%%%%%%%%%%%%%%%%%%%%%%%%%%%%%%%%%%%%%%%%%%%%%%%%%%%%%%%%%%

%%%%%%%%%%%%%%%%%%%%%%%%%%%%%%%%%%%%%%%%%%%%%%%%%%%%%%%%%%
%%%%%%%%%%%%%%%%%%%%%%%%%%%%%%%%%%%%%%%%%%%%%%%%%%%%%%%%

\begin{figure}[th]
\centerline{
\includegraphics[height=0.38\textwidth, angle=90]{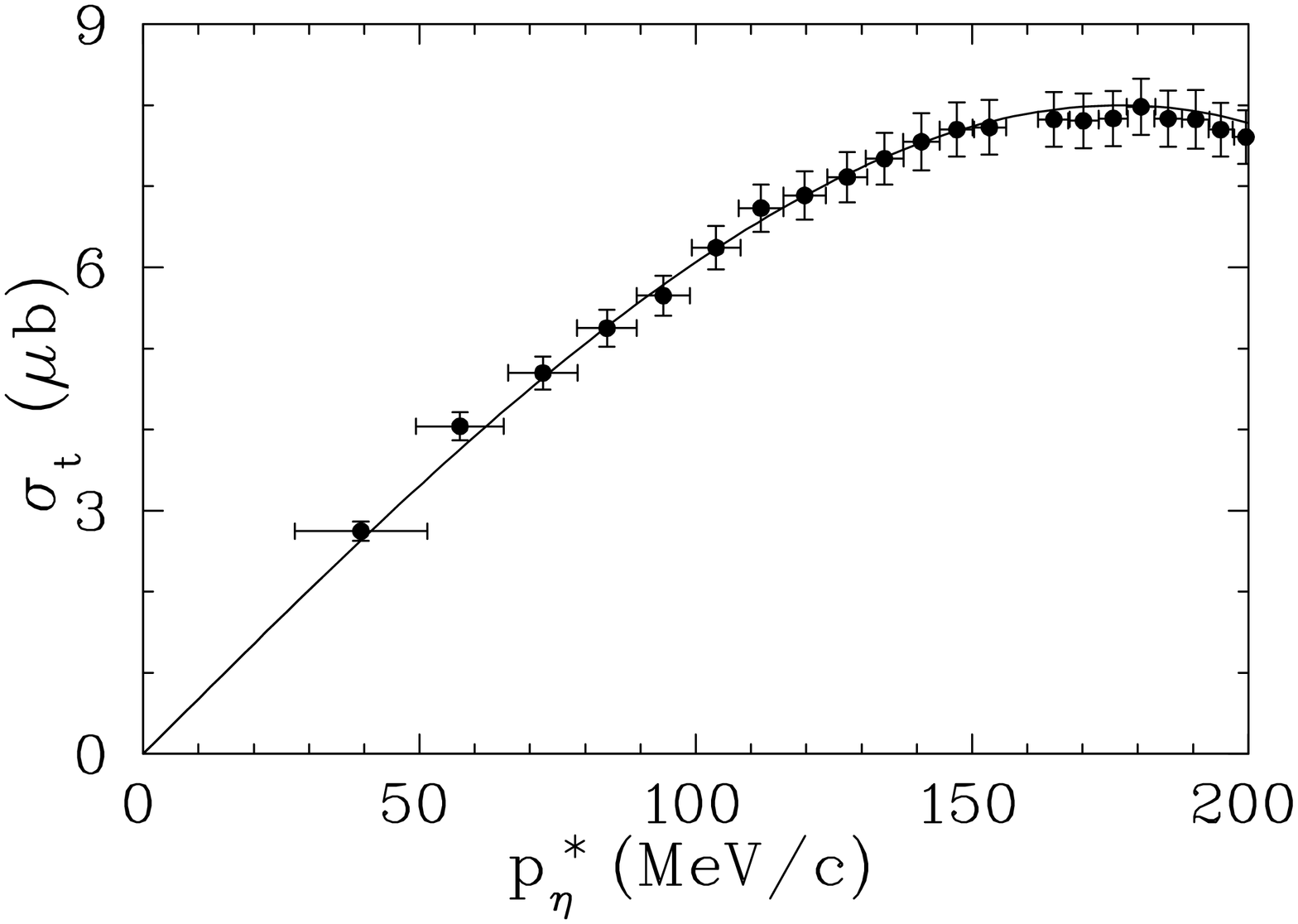}}
\caption{Our total cross section (circles) for
	$\gamma p\to\eta p$ as a function of the
	$\eta$ momentum in the c.m. frame.  The
	solid line shows the results of fitting 
	our data to a sum of linear and cubic 
	terms. \label{fig:g6}}
\end{figure}

Traditionally, to illustrate resonance masses and widths,
the total cross section is plotted
as a function of the c.m. energy.  As seen in 
Figs.~\ref{fig:g2} and \ref{fig:g2a}, the $\gamma p\to\eta 
p$ total cross section rises sharply above the reaction 
threshold.  Such behavior is usually attributed to the 
dominance of the $N(1535)1/2^-$ resonance, having a mass 
close to the $\eta$ production threshold ($W = 1487$~MeV) 
and a strong coupling to the $\eta N$ channel. Generally, 
the cross section for any process with a two-particle final 
state has the form [(p*/W)~F(W)]. The first factor comes 
from  the phase-space integration, and $p^\ast$ is the 
final-state relative momentum in the c.m. frame. The 
second factor F(W) is determined by amplitudes. The essential 
point is that W and F(W) depend explicitly only on 
$(p^\ast)^2$, not on $p^\ast$. In terms of the 
final-state parameters, $W$ depends on masses and 
$(p^\ast)^2$. Therefore, the near-threshold structure 
of the cross section should look as a series in the odd 
powers of $p^\ast$.  Our data for $\gamma p\to\eta p$ 
are well described up to $p^\ast_\eta\sim 200$~MeV/c as 
$a_1 p^\ast_\eta + a_3 (p^\ast_\eta)^3$ with $a_1 = 
(6.79\pm 0.09)\times 10^{-2}~\mu b$/(MeV/c) and $a_3 = 
-(7.24\pm0.22)\times 10^{-7}~\mu b$/(MeV/c)$^3$ 
(see Fig.~\ref{fig:g6}). Contributions to the cubic 
term come both from P-wave amplitudes and from 
$W$-dependence of the S-wave amplitude, which is 
essential due to the near-theshold dominance of the 
S-wave resonance $N(1535)1/2^-$. Note that the 
characteristic momentum for changes in our fit is 
$\sqrt{|a_1/a_3|}\sim 300$~MeV/c, while the 
maximum of the resonance peak corresponds to 
$p^\ast_\eta\sim 175$~MeV/c (see Fig.~\ref{fig:g6}).
The good quality of our data reveals itself in very 
small fluctuations of experimental points with 
respect to the fit. For comparison, a similar 
threshold behavior has been also observed in
$\pi^-p\to\eta n$~\cite{ar05}, but the 
fluctuations there were larger due to lower 
precision of the data, and the coefficient $a_3$ 
could not be determined reliably. In addition, 
our fit gives implicit confirmation of small 
coupling of the $\eta N$ channel with the $D$-wave 
resonance $N(1520)3/2^-$, which could generate 
an essential term $(p^\ast_\eta)^5$.
%%%%%%%%%%%%%%%%%%%%%%%%%%%%%%%%%%%%%%%%%%%%%%%%%%%%%%%
%%%%%%%%%%%%%%%%%%%%%%%%%%%%%%%%%%%%%%%%%%%%%%%%%%%%%%%

\begin{figure}[th]
\centerline{
\includegraphics[height=0.38\textwidth, angle=90]{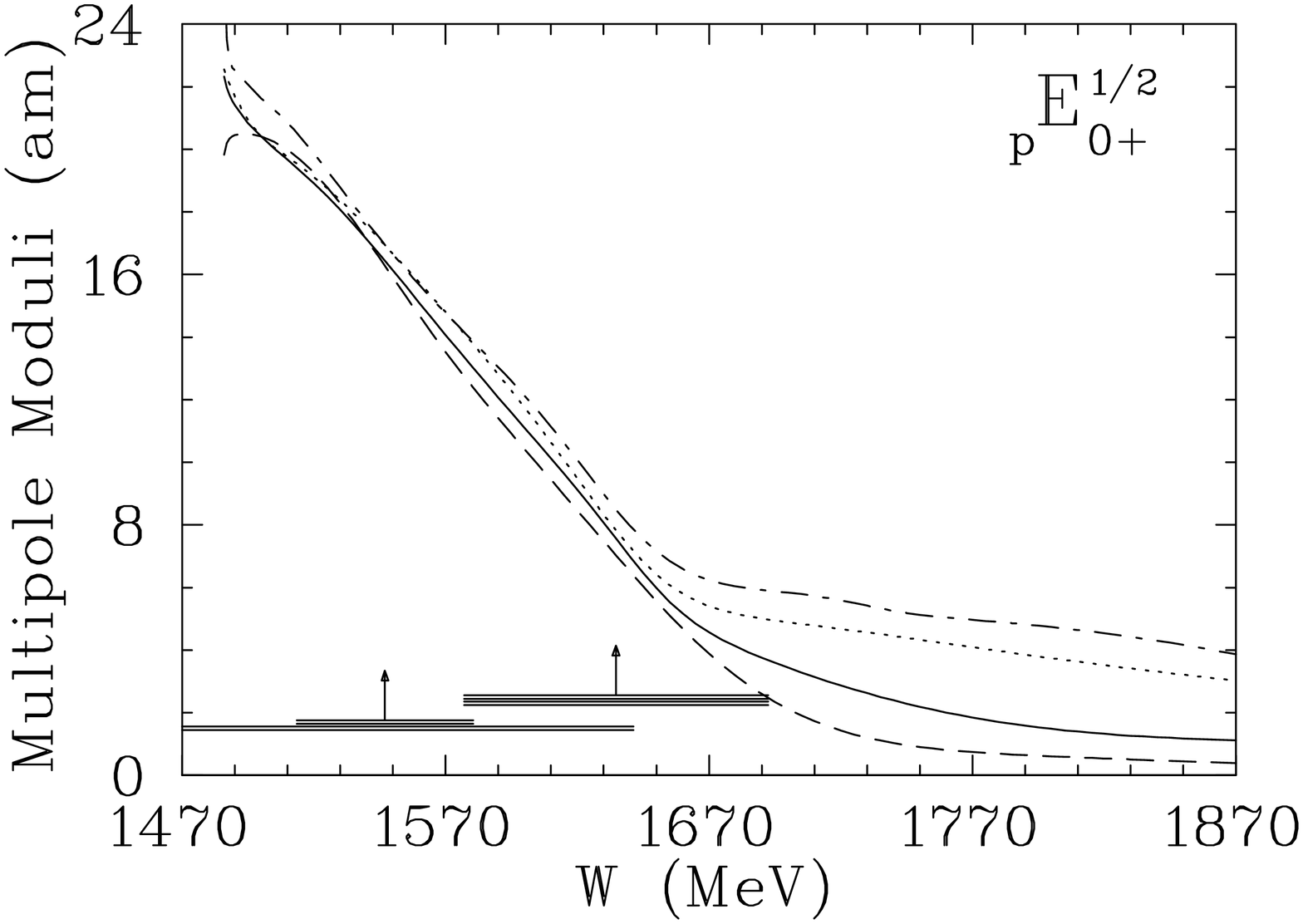}}
\caption{Modulus of the multipole amplitude S$_{11}$pE
    ($_pE^{1/2}_{0+}$) for $\gamma p\to\eta p$ from 
    the reaction threshold to $E_{\gamma}=1.4$~GeV.  
    Notations for the amplitude curves are the same 
    as in Fig.~\protect\ref{fig:g3}.  The vertical 
    arrows indicate $W_R$ (Breit-Wigner mass) and 
    the horizontal bars show the full and partial 
    width $\Gamma$ for $\Gamma_{\pi N}$ 
    associated with the SAID solution SP06 for 
    $\pi N$~\protect\cite{sp06}. \label{fig:g7}}
\end{figure}

Due to the dominance of the low-energy $S$-wave multipole,
this contribution is nearly model-independent up to 
$W=1650$~MeV. The modulus of the corresponding 
amplitude is plotted in Fig.~\ref{fig:g7} for both 
the SAID and MAID solutions. Phase differences are 
possible - these can be resolved in coupled-channel 
fits~\cite{Paris}.  Figure~\ref{fig:g7} also shows 
the Breit-Wigner parameters, masses and widths, of 
two $S_{11}$ resonances as found in the SAID PWA 
solution SP06 for the $\pi N$ elastic 
scattering~\cite{sp06}. Note that $N(1650)1/2^-$ seems 
to be purely elastic, i.e., coupled only to the $\pi N$ 
channel. If so, its contribution to $\eta$ 
photoproduction should be small.
%%%%%%%%%%%%%%%%%%%%%%%%%%%%%%%%%%%%%%%%%%%%%%%%%%%%%%%
%%%%%%%%%%%%%%%%%%%%%%%%%%%%%%%%%%%%%%%%%%%%%%%%%%%%%%%

%%%%%%%%%%%%%%%%%%%%%%%%%%%%%%%%%%%%%%
\section{Summary and conclusions}
\label{sec:Conc}

The $\gamma p\to\eta p$ differential cross sections 
have been measured at the tagged photon facility
of the Mainz Microtron MAMI-C using the Crystal Ball
and TAPS spectrometers. The data span the photon-energy
range 707~MeV to 1402~MeV and the full angular range
in the c.m. frame. The accumulation of $3\times 10^6$
$\gamma p\to\eta p\to 3\pi^0p\to 6\gamma p$ events 
allows the fine binning of the data in energy and angle,
which will enable the reaction dynamics to be studied
in greater detail than previously possible. The 
present data agree well with previous equivalent
measurements, but are markedly superior in terms of
precision and energy resolution.

The present cross sections for the free proton show no 
evidence of enhancement in the region $W\sim1680$~MeV, 
contrary to recent equivalent measurements on the 
quasi-free neutron~\cite{graal,CB-Elsa,Lns}.
However, this does not exclude the existence of
an $N^\ast(1680)$ state as hypothesized in Ref.~\cite{p11}.
In the region around $W=1680$~MeV, 
we rather observe a dip structure that becomes more 
pronounced at forward production angles of $\eta$.  
This feature was missed or questionable in the analysis 
of the previous data. The interpretation of this dip 
depends on dynamics.

Our $\gamma p\to\eta p$ data points have been included 
in a new SAID (GE09) and Regge-MAID PWAs, to which 
they made an substantial contribution, particularly 
for forward angles. Comparing to the previous SAID fit, 
E429, and to the $\eta$-MAID fit, the description of all 
existing data by the new solutions, GE09 and Regge-MAID, 
is more satisfactory in the entire energy range.

We expect that the data presented in this paper will be 
invaluable for future partial-wave and coupled-channel 
analyses, in that they can provide much stronger 
constraints on the properties of the nucleon resonances 
from our energy region.

%%%%%%%%%%%%%%%%%%%%%%%%%%%%%%%%%%%%%%%%%%%%%%%%%%
\acknowledgments

The authors thank the accelerator group of MAMI for
their support.  We thank the undergraduate students
of Mount Allison University and The George Washington 
University for their assistance. This work was 
supported by the U.S. DOE (Grants No. DE--FG02--99ER41110 
and DE--FG02--01ER41194), NSF (Grant No. PHY 0652549), 
the Jefferson Laboratory, the Southeastern Universities
Research Association (DOE Contract No.
DE--AC05--84ER40150), the U.K. Engineering and Physical
Sciences Research Council, the U.K. Science and Technology
Facilities Council, the Deutsche Forschungsgemeinschaft
(SFB 443,SFB/TR 16), the European Community-Research
Infrastructure Activity under the FP6 ``Structuring
the European Research Area" program (Hadron Physics,
Contract No. RII3--CT--2004--506078), DFG-RFBR (Grant
No. 09--02--91330) of Germany and Russia, SNF of
Switzerland, NSERC of Canada, and Russian State Grant
SS--3628.2008.2.

%%%%%%%%%%%%%%%%%%%%%%%%%%%%%%%%%%%%%%%%%%%%%%%%%%%%%

%%%%%%%%%%%%%%%%%%%%%%%%%%%%%%%%%%%%%%%%%%%%%%%%%%%%%%%%
\end{document}